\documentstyle[preprint,aps]{revtex}
\begin{document}
\preprint{\begin{minipage}{2in}\begin{flushright}
  SLAC-PUB-7121 (T) \\[-3mm] hep-lat/9603016 \end{flushright}
  \end{minipage}}
\input epsf
\draft
\title{CORE Technology and Exact Hamiltonian Real-Space
Renormalization Group Transformations}
\author{Colin~J.~Morningstar}
\address{Dept.~of Physics, University of California at San Diego,
  La Jolla, California 92093-0319}
\author{Marvin Weinstein}
\address{Stanford Linear Accelerator Center, Stanford University,
  Stanford, California 94309}
\date{March 20, 1996}
\maketitle
\begin{abstract}
The COntractor REnormalization group (CORE) method, a new approach to
solving Hamiltonian lattice systems, is presented.  The method
defines a systematic and nonperturbative means of implementing
Kadanoff-Wilson real-space renormalization group transformations
using cluster expansion and contraction techniques.  We illustrate
the approach and demonstrate its effectiveness using scalar field
theory, the Heisenberg antiferromagnetic chain, and the anisotropic
Ising chain.  Future applications to the Hubbard and t-J models and
lattice gauge theory are discussed.
\end{abstract}
%

\section{Introduction}

Whether we wish to compute the mass spectrum of lattice QCD or the
phase structure of the extended Hubbard model, we are faced with the
same problem---extracting physics from a theory to which conventional
perturbative methods cannot be applied.  To date, the most popular
approach to these problems has been Monte Carlo evaluation of the
Feynman path integral.  Recently, we introduced an alternative,
Hamiltonian-based approach called the CORE (Contractor Renormalization
Group) approximation\cite{COREPrl}
and applied it to the case of the 1+1-dimensional
Ising model.  In this paper, we significantly extend the method and
simplify its implementation.  The CORE approach defines a
systematic, nonperturbative, and computable means of carrying out a
Hamiltonian version of the Kadanoff-Wilson\cite{KadWil} real-space
renormalization group transformation for lattice field theories and
lattice spin systems.  The method relies on contraction and
cluster expansion techniques.

The CORE approximation improves upon other methods of implementing
approximate real-space renormalization group transformations on
Hamiltonian systems\cite{whiteetal} in several ways. First, our methods
make it possible to define a {\em gauge-invariant} renormalization
group transformation for any abelian or nonabelian lattice gauge
theory, something which was not possible in earlier schemes.  Second,
it is no more difficult to treat fermions than bosons when one uses
these methods.  Third, it is easy to add a chemical potential to the
Hamiltonian for a system such as the Hubbard model in order to tune the
density of the ground state, a difficult feat in earlier Hamiltonian
real-space renormalization group schemes.  Finally, CORE allows us to
map a theory with one set of degrees of freedom into a theory described
in terms of a very different set of degrees of freedom but possessing
the same low-energy physics.  Within the context of lattice gauge
theories, this means we can start from a theory of quarks and gluons
and map it into a system in which the effective degrees of freedom have
the quantum numbers of mesons and baryons.  Computing such a
transformation within the Hamiltonian framework was not possible in
earlier methods.

In addition to the above qualitative improvements, there are also
substantial quantitative refinements.  For example, earlier attempts to
compute the ground-state energy density and other properties of the
$1+1$-dimensional Heisenberg antiferromagnet using previous real-space
renormalization group methods\cite{HAFrsrg} or $t$-expansion
techniques\cite{texpansion} had difficulty matching the accuracy of
Anderson's\cite{Anderson} naive spin-wave approximation.  We will
demonstrate that the CORE approximation significantly improves on
Anderson's calculation without making any large spin approximations.
Another example which we discuss is the $1+1$-dimensional Ising model.
We will show that an easily implemented CORE computation substantially
improves upon results from earlier methods.

We close this section with a brief review of the renormalization group
(RG) in order to contrast the CORE method to previous RG implementations.
Next, in Sec.~\ref{sec:rules}, we state without proof
the rules for carrying out a CORE calculation. We then illustrate the
method in Sec.~\ref{sec:examples} by applying the rules in four
examples: free scalar field theory with single-state truncation, the
Heisenberg antiferromagnetic spin chain with two-state truncation, the
anisotropic Ising model with two-state truncation, and free scalar
field theory with an infinite-state truncation scheme.  The rules are
then derived in Sec.~\ref{sec:derive}.  In Sec.~\ref{sec:converge}, two
issues are discussed:  the use of approximate contractors, to
establish the connection of the CORE approach to earlier methods; and
the convergence of the cluster expansion, to demonstrate the
need for summation via the renormalization group.
Finally, future applications to the Hubbard and extended Hubbard models
and lattice gauge theory with and without fermions are discussed in
Sec.~\ref{sec:looking}.  We also address the issue of relating the
contractor renormalization group to the familiar perturbative
renormalization group in $\phi^4$ theory.

\subsection{Preliminary Remarks}

Physical systems in quantum field theory and statistical mechanics
involve a large number of degrees of freedom and can usually be
described in terms of a local Hamiltonian.  Conventional wisdom says
that when the coherence length of such a system is small, the
properties of the system depend strongly on the form and strengths of
the interactions in the Hamiltonian; whereas, when the coherence
length is large, many degrees of freedom behave cooperatively and the
properties of the system are governed primarily by the nature of this
cooperation with the detailed form of the Hamiltonian playing only a
subsidiary role.

The renormalization group\cite{KadWil}, as formulated by Kadanoff and
Wilson, is generally thought of as a method for treating systems in
which the coherence length encompasses many degrees of freedom.  This
method is based on iteratively {\it thinning} the degrees of freedom in
the problem, an approach which is similar to that followed in
hydrodynamics wherein the innumerable microscopic degrees of freedom
are replaced by a much smaller set of spatially-averaged, macroscopic
variables, such as the density and pressure. In this renormalization
group method, the thinning is achieved via a sequence of
renormalization group transformations.

While the original formulation of the RG method was done for the
partition function or its path-integral analogue in field theory, the
approach has been extended to Hamiltonian systems.  The basic idea is
to construct a real-space renormalization group transformation, $\tau$,
which maps the Hamiltonian $H_0$ of a theory defined on some lattice
$L_0$ to a new theory $H_1$ defined on a {\em coarser} lattice $L_1$ in
such a way that the new theory has the {\em same low-energy physics} as
the original theory.  To extract the low energy physics of the original
theory, we repeatedly apply the transformation $\tau$ and generate the
sequence of {\em renormalized Hamiltonians:} $H_1=\tau(H_0)$,
$H_2=\tau(H_1)$, $H_3=\tau(H_2)$, $\dots$. This sequence usually
approaches a fixed point of $\tau$, that is, a Hamiltonian $H^\ast$
satisfying $\tau(H^\ast)=H^\ast$. Each renormalized Hamiltonian in this
sequence possesses the {\it same} low-energy physics, but the degrees
of freedom have been {\it thinned}. Eventually, the number of remaining
degrees of freedom lying within the coherence length will be small and
the resulting Hamiltonian will be more amenable to solution.

Generally, the same transformation $\tau$ is used for each iteration;
however, this is not required. The use of different transformations for
each iteration is clearly impractical, but the use of a different
transformation for the first one or few steps could be a powerful
generalization of the method, facilitating great simplifications.  A
quantum field theory could be mapped into a generalized spin model; QCD
could be mapped into a theory of interacting hadrons.

Defining and carrying out the thinning transformations is the key to
the RG approach.  The RG transformation $\tau$ is usually defined by
requiring invariance of the partition function or its path-integral
analogue in field theory.  The RG method exactly describes the
low-lying physics as long as $\tau$ can be exactly implemented, which
is rarely the case.  In practice, approximations must be made, such as
those made in the $\epsilon$-expansion \cite{epsilon}, the use of
perturbative matching as in the heavy-quark effective field
theory\cite{hqet} and nonrelativistic QCD\cite{nrqcd}, and stochastic
estimation as in the Monte Carlo renormalization group\cite{mcrg}
approach.  The CORE approach is a new and powerful method for defining
$\tau$ and computing $\tau(H)$ which relies on contraction and cluster
techniques. In contrast to other methods, the approximations made in
the CORE approach do not limit the usefulness of the method to
any restricted range of coupling constants or other parameters in the
theory.  The CORE approach works well not only near a critical point
when the coherence length is large, but also in instances where it is
small.  It is a general method for solving any lattice Hamiltonian
problem.

CORE computations begin by defining the way in which the new lattice is
{\em coarser} than the original lattice.  We begin by partitioning the
lattice into identical blocks.  The Hilbert space of states
corresponding to each block is then {\em truncated} by discarding all
but a certain number of low-lying states; we generally retain enough
states so that the truncated degrees of freedom on a block resemble
those of a site on the original lattice. The {\em renormalized} or
effective Hamiltonian $H^{\rm ren}$ in this truncated space of states
is then defined in terms of the original Hamiltonian $H$ by
\begin{equation}
 H^{\rm ren}= \lim_{t\rightarrow\infty}\,\,
 [ \! [\, T(t)^2 \,] \! ]^{-\frac{1}{2}}\ [ \! [ T(t)\ H
  \ T(t) \,] \! ]\ [ \! [\, T(t)^2 \,] \! ]^{-\frac{1}{2}},
 \label{hamren}
\end{equation}
where $T(t)=e^{-tH}$ is the {\em contractor} and $[\! [\dots ]\! ]$
refers to truncation to the subspace of retained states.
There is a one-to-one correspondence between the eigenvalues of the
renormalized Hamiltonian and the low-lying eigenvalues of the
original Hamiltonian.  In general, the renormalized Hamiltonian
cannot be exactly determined; CORE approximates $H^{\rm ren}$ using
a finite cluster expansion, an approximation which can be
systematically improved.  Matrix elements of various operators can
also be evaluated in CORE by defining a sequence of renormalized
operators.

We use the phrase ``CORE technology'' to refer to the set of tools
which allow us to systematically and nonperturbatively compute an
arbitrarily-accurate approximation to the exact renormalization group
transformation for a lattice field theory or spin system without having
to diagonalize the original infinite-volume theory.  The power of these
methods is that usually only a few terms in the cluster expansion of
the renormalized Hamiltonian yield remarkably good results.

\section{The Rules}
\label{sec:rules}

In this section, we state, without proof, the rules for carrying out a
CORE computation.  We assume that we are studying a theory defined by
a local Hamiltonian $H$ on a regular lattice of infinite extent in
some number of dimensions.

A CORE computation proceeds as follows:

\begin{enumerate}
\item    First, divide the lattice into identical,
         disjoint blocks ${\rm B}_j$.  Denote the space of states
         associated with block ${\rm B}_j$ by ${\cal H}_j$ and denote
         the common dimension of each of these spaces by $N$.

\item    Define a truncation scheme by selecting a low-lying subspace
         ${\cal P}_j \subset{\cal H}_j$ of dimension $M < N$ on every
         block; the same subspace should be chosen on each block.
         In what follows, we will denote the $M$ retained states by
         $\vert\phi_s(j)\rangle$ and use them to construct the
         projection operators
         \begin{eqnarray}
               P(j) &=& \sum_{s=1}^M \vert\phi_s(j)\rangle
               \langle\phi_s(j)\vert, \\
               P    &=& \prod_j P(j).
            \label{introproj}
         \end{eqnarray}
         Let $[ \! [\dots] \! ]$ denote truncation to the subspace
         spanned by the taking tensor products of the states
         $\vert\phi_s(j)\rangle$.  Thus, for any operator $O$, the
         truncated operator is defined as $[ \! [ O ] \! ] = P^\dagger OP$.
         Note, choosing to retain states such that the truncated
         degrees of freedom on a single block resemble
         those associated with a single site on the original lattice
         ensures that the renormalized Hamiltonian will take
         a form similar to that of the original Hamiltonian, facilitating
         the iteration process; however, sometimes it is useful to make
         a different choice and map the original theory into one
         formulated in terms of new degrees of freedom.

\item    Compute (see below) the renormalized Hamiltonian defined in
         Eq.~\ref{hamren}, $H^{\rm ren}_1=\tau(H)$, and the renormalized
         operators corresponding to any matrix elements of interest.

\item    Repeat the above steps using $H^{\rm ren}_m$ to obtain
         $H^{\rm ren}_{m+1}=\tau(H^{\rm ren}_m)$.  Iterate this
         process until the renormalized Hamiltonian is simple
         enough that its low-lying eigenvalues can be
         determined.
\end{enumerate}

Because the Hamiltonian is {\it extensive} (a concept we will define
later) and the block-by-block truncation preserves this property, the
renormalized Hamiltonian can be approximated using the finite cluster
method (FCM).  This method was first used by Domb\cite{Domb} in the
application of the Mayer cluster integral theory to the Ising model.  A
formal proof of the method in the Ising and Heisenberg models was then
presented by Rushbrooke\cite{Rushbrooke}.  The method was later
generalized by Sykes {\it et al.}\cite{Sykes}.  The finite cluster
method expresses any extensive quantity in an infinite volume as a sum
of finite-volume contributions.  The procedure is simple to implement
and provides numerous means of detecting computational errors.  A
general statement of the method can be found in Ref.~\cite{DombB}.

Evaluation of $H^{\rm ren}$ by the finite cluster method is
accomplished in the following sequence of steps:

\begin{enumerate}
\item  Compute the renormalized Hamiltonian for a theory
       defined on a sublattice which contains only a single
       block $B_j$ (how this is done will be described below).  Denote
       this Hamiltonian
       by $H^{\rm r}(B_j)=h_1(B_j)$.  This yields all of the
       so-called range-1 terms in the cluster expansion of the
       renormalized Hamiltonian.

\item  Calculate the renormalized Hamiltonian
       $H^{\rm r}(B_j,B_{j+1})$ for a
       theory defined on a sublattice made up of two adjacent
       ({\it connected}) blocks $B_j$ and $B_{j+1}$.  The range-2
       contributions to the cluster expansion of the renormalized
       Hamiltonian on the infinite lattice are obtained by
       removing from $H^{\rm r}(B_j,B_{j+1})$ those contributions which
       arise from terms already included in the single block
       calculation:
         \begin{equation}
            h_2(B_j,B_{j+1}) = H^{\rm r}(B_j,B_{j+1})-h_1(B_j)
                      - h_1(B_{j+1}).
          \label{Hrangetwo}
         \end{equation}

\item  Repeat this procedure for sublattices containing
       successively more connected blocks.  For example, for a
       sublattice consisting of three adjacent blocks $B_j$,
       $B_{j+1}$, and $B_{j+2}$, use
         \begin{eqnarray}
            h_3(B_j,B_{j+1},B_{j+2}) &=& H^{\rm r}(B_j,B_{j+1},B_{j+2})
               -h_2(B_j,B_{j+1})- h_2(B_{j+1},B_{j+2})\nonumber\\
             &-&h_1(B_j)- h_1(B_{j+1}) - h_1(B_{j+2}).
          \label{Hrangethree}
         \end{eqnarray}
       Since the renormalized Hamiltonian is extensive, only connected
       sublattices need to be considered.  Recall that a quantity
       is extensive if, when evaluated on a disconnected sublattice,
       it is the sum of that quantity evaluated separately on the
       connected components of the sublattice.  A truncated cluster
       expansion can then be defined by neglecting clusters larger
       than some specified range.

\item  To complete the determination of the renormalized Hamiltonian
       $H^{\rm ren}$ on the infinite lattice, sum the connected
       contributions $h_r(B_j,\dots)$ from the finite sublattices
       according to their embeddings in the full lattice.
       For example, on an infinite one-dimensional lattice:
         \begin{equation}
           H^{\rm ren} = \sum_{j=-\infty}^{\infty}
             \sum_{r=1}^\infty h_r(B_j,\dots, B_{j+r-1}).
         \end{equation}
       Express $H^{\rm ren}$ in terms of block-variables such that
       the form of $H^{\rm ren}$ resembles that of the previous
       Hamiltonian in the RG sequence.
\end{enumerate}

The key ingredient of CORE is the method used to explicitly construct
the renormalized Hamiltonian $H^{\rm r}(G)$ and other renormalized
operators on a given cluster or sublattice $G$.  While, in principle,
the appropriate generalization of Eq.~\ref{hamren} completely specifies
what has to be done, in practice, an attempt to compute this quantity
by brute force will run into problems since the operator $[[ T(t) ]]$
becomes singular as $t \rightarrow \infty$.  To see how this problem
arises, consider a sublattice comprised of $R$ connected blocks $B_1
\dots B_R$.  Let $H(G)$ denote the Hamiltonian obtained by restricting
the infinite lattice $H$ to the sublattice $G$ and suppose that we
truncate to the subspace ${\cal P}={\cal P}_1\otimes\cdots\otimes{\cal
P}_R$ spanned by the $M^R$ states $\vert\phi_\alpha(G)\rangle$.
Remember that the states $\vert\phi_\alpha(G)\rangle$ are tensor
products of the retained states on each of the $R$ blocks in the
cluster $G$.  Let us denote by $\vert\epsilon_\beta(G)\rangle$ the
eigenstates of $H(G)$ with eigenvalues $\epsilon_\beta(G)$ and expand
the states $\vert\phi_\alpha(G)\rangle$ in terms of these eigenstates:
{\it i.e.,}
\begin{equation}
   \vert\phi_\alpha(G)\rangle =
      \sum_\beta \alpha_\beta \vert\epsilon_\beta(G)\rangle.
   \label{stateexpansion}
\end{equation}
It then follows that
\begin{equation}
   T(t) \vert\phi_\alpha(G)\rangle =
      \sum_\beta \alpha_\beta
      e^{-t\epsilon_\beta(G)}\,\vert\epsilon_\beta(G)\rangle ,
   \label{stateevolve}
\end{equation}
from which we see that all states $\vert\phi_\alpha(G)\rangle$ which
have a nonvanishing overlap with the ground state of $H(G)$ contract
onto the same state as $t \rightarrow \infty$.  This causes great
difficulties if we attempt to numerically compute $[[ T(t)^2]]^{-1/2}$.
Fortunately, there is an elegant and simple solution to this
problem which avoids explicit computation of $[[ T(t)^2 ]]^{-1/2}$:
make a unitary (or orthogonal) change of basis,
$S(G)$, on the states $\vert\phi_\alpha(G)\rangle$ such
that each state in the new basis contracts onto a unique eigenstate of
$H(G)$.  In this new basis, the computation of $H^{\rm r}(G)$ is then
straightforward. The discussion which follows specifies the rules for
computing the necessary change of basis and $H^{\rm r}(G)$ for a
general $H(G)$. We state these rules in full generality so as to
allow for the special situation in which $H(G)$ has degenerate
eigenvalues, and then apply them to successively more complicated
examples in order to show how they work in practice.

$H^{\rm r}(G)$ and the change of basis $S(G)$ may be determined as
follows:

\begin{enumerate}
\item  Find the eigenstates $\vert\epsilon_\beta(G)\rangle$
       and corresponding eigenvalues $\epsilon_\beta(G)$ of
       $H(G)$, where $\beta=0\dots N^R-1$.  Order these states
       so that $\epsilon_\beta \leq \epsilon_{\beta+1}$.

\item  Construct the $M^R\times N^R$ matrix
       \begin{equation}
         Q(G)_{\alpha\beta} = \langle\phi_\alpha(G)\vert
            \epsilon_\beta(G)\rangle.
       \end{equation}
       Each row of $Q(G)$ gives the expansion of one of the
       retained states in terms of the eigenstates of $H(G)$.
       Each column of $Q(G)$ gives the projection of some
       eigenstate into the truncated subspace.  Also, let
       $S(G)$ be the $M^R\times M^R$ identity matrix.
       Set $m=M^R$, $p=0$, and $q=0$.

\item\label{stepa}
       Copy the first $g$ columns of $Q(G)$ into an $m\times g$
       matrix $C$, where $g$ is the degeneracy of the lowest-lying
       eigenvalue.  If the ground state of the cluster is nondegenerate,
       then $g=1$.  The columns of $C$ correspond to
       the degenerate eigenstates $\vert\epsilon_q(G)
       \rangle, \cdots, \vert\epsilon_{q+g-1}(G)\rangle$.
       Having formed $C$, perform a singular value
       decomposition (SVD), writing
       \begin{equation}
           C = U\ \Sigma\ V^\dagger,
       \end{equation}
       where $U$ is an $m\times m$ unitary matrix,
       $V$ is a $g\times g$ unitary matrix, and
       $\Sigma$ is an $m\times g$ matrix of the form
       \begin{eqnarray}
           \Sigma &=& \left(\begin{array}{ll}
             \Delta_{r\times r} &
             {\bf 0}_{r\times (g-r)} \\ {\bf 0}_{(m-r)\times r} &
             {\bf 0}_{(m-r)\times (g-r)} \end{array}\right), \\[2mm]
           \Delta &=& {\rm diag}(\sigma_1, \cdots,
             \sigma_r),
       \end{eqnarray}
       where the elements $\sigma_j$ are real and satisfy
       $\sigma_1\geq \sigma_2\geq \dots \geq \sigma_r > 0$
       and $r\leq {\rm min}(m,g)$ is the {\em rank} of the
       matrix $C$.  In other words, use the SVD\cite{svd} to
       construct orthonormal bases for the nullspace and range
       of the matrix $C$.  Note that the SVD theorem guarantees
       that such a decomposition exists and that $\Sigma$
       is unique.

\item  Multiply $U^\dagger\ Q(G)$ and, by abuse of notation, once again
       call the result $Q(G)$.
       Then discard the first $g$ columns and the first
       $r$ rows of the new $Q(G)$.  The resulting matrix, which we
       again call $Q(G)$, is now an $(m-r) \times (N^R-q-g)$ matrix.
       Note, $r$ may be zero.

\item\label{stepb}
       Form the matrix
       \begin{equation}
          R = \left(\begin{array}{cc} {\bf 1}_{p\times p} &
              {\bf 0}_{p\times m} \\
              {\bf 0}_{m\times p} &
              U^\dagger_{m\times m}\end{array}\right),
       \end{equation}
       and multiply $R\ S(G)$; call the result $S(G)$.
       Define the states
       \begin{equation}
          \vert{\cal T}_{p+s-1}(G)\rangle =  \sum_{s^\prime=1}^g
           \ V_{s^\prime s}\ \vert\epsilon_{q+s^\prime-1}(G)\rangle,
       \label{calTdef}
       \end{equation}
       with corresponding degenerate eigenvalues ${\cal T}_{p+s-1}(G)
       = \epsilon_{q}(G)$, for $s=1\dots r$.  Set $p\rightarrow p+r$,
       $q\rightarrow q+g$, and $m\rightarrow m-r$.

\item  Repeat steps \ref{stepa} to \ref{stepb} with higher
       and higher energy eigenvalues until $p=M^R$.  In step
       \ref{stepa}, $g$ is now the degeneracy of the lowest-lying
       {\em remaining} eigenvalue.  At the end of this procedure,
       we will have constructed a unitary $M^R\times M^R$ matrix $S(G)$
       and a set of eigenstates $\{\vert{\cal T}_\beta(G)\rangle\}$
       with energy eigenvalues ${\cal T}_\beta(G)$ for $\beta=1
       \dots M^R$.

       In the discussion which follows, it will be convenient
       to make the following definitions:

\begin{quote}
{\bf Definition:}  \label{remnantdef}
       The eigenstates $\vert{\cal T}_\beta(G)\rangle$
       are referred to as the {\em remnant eigenstates} of $H(G)$
       in ${\cal P}$.  The set of these $M^R$ remnant eigenstates is
       called the {\em contraction remnant}.  The matrix $S(G)$ is
       referred to as the {\em triangulation matrix}.
\end{quote}

       As we already noted, the triangulation matrix $S(G)$ is
       simply a change of basis, taking us from the original basis
       $\{\vert\phi_\alpha(G)\rangle\}$ of retained
       tensor-product states in the truncated subspace to
       another basis $\{\vert\xi_\alpha(G)\rangle\}$
       in which only the first state has a nonvanishing overlap
       with the ground eigenstate, only the first and second
       states have nonzero overlaps with the first excited
       eigenstate, and so on; hence, $S(G)_{\alpha\beta}
       =\langle\xi_\alpha(G)\vert\phi_\beta(G)\rangle$.
       The remnant eigenstates are
       essentially the $M^R$ lowest-lying eigenstates of $H(G)$
       whose projections into ${\cal P}$ are nonvanishing and cannot
       be written as linear combinations of lower-energy eigenstates
       projected into ${\cal P}$.  In other words, the projections
       of the remnant eigenstates into ${\cal P}$ are all linearly
       independent.  Within degeneracy subspaces, the eigenstates
       must be rotated in order to eliminate all linear combinations
       whose projections in ${\cal P}$ are zero or completely
       expressible in terms of the projections of lower-lying
       eigenstates.  Note that the singular value decomposition
       theorem \cite{svd} guarantees the existence of the
       triangulation matrix and the contraction remnant.

\item  In the basis of the remnant eigenstates, construct the matrices
       \begin{eqnarray}
          H_{\cal T}(G)_{\alpha\beta} &=& \langle{\cal T}_\alpha(G)\vert
            H(G)\vert{\cal T}_\beta(G)\rangle = \delta_{\alpha\beta}
            {\cal T}_\alpha(G),\\
          O_{\cal T}(G)_{\alpha\beta} &=& \langle{\cal T}_\alpha(G)\vert
            O(G)\vert{\cal T}_\beta(G)\rangle,
       \end{eqnarray}
       where $O(G)$ is some operator of interest defined on the
       sublattice $G$.

\item  The renormalized operators are at last
       given in terms of the triangulation matrix and the
       operators evaluated in the contraction remnant:
       \begin{eqnarray}
           H^{\rm r}(G) &=& S^\dagger(G)\ H_{\cal T}(G)\ S(G),\\
           O^{\rm r}(G) &=& S^\dagger(G)\ O_{\cal T}(G)\ S(G).
       \end{eqnarray}
\end{enumerate}

Note that the CORE approach described here differs from that described
previously\cite{COREPrl}.  In our earlier formulation of the method,
the contractor $T(t)$ in Eq.~\ref{hamren} was approximated by a product
of exactly computable exponentials.  The variable $t$ was then treated
as a variational parameter, adjusted so as to minimize the mean-field
energy in each RG iteration.

\section{Four Examples}
\label{sec:examples}

To better illustrate the method and demonstrate its effectiveness,
we now apply these rules in four examples.  Each of these
examples, free scalar field theory with single-state truncation,
the $1+1$-dimensional Heisenberg antiferromagnet with two-state
truncation, the $1+1$-dimensional Ising model with two-state
truncation, and free scalar field theory with infinite-state
truncation, has been chosen to clarify a particular aspect
of the rules.

\subsection{Single-State Truncation: Free Scalar Field Theory}

First let us discuss a massless $(\mu=0)$
free-field theory.  Free scalar field
theory on a lattice is just a set of coupled harmonic oscillators,
 \begin{equation}
      H = \sum_j \left[ \frac{1}{2} \Pi(j)^2 + \frac{\mu^2}{2} \phi(j)^2
          + \frac{1}{2} ( \phi(j+1) - \phi(j) )^2 \right] ,
   \label{bosefreefield}
 \end{equation}
where $[ \phi(j), \Pi(k) ] = i \delta_{jk}$.
The simplest possible truncation procedure we can adopt is to keep
the number of sites fixed and truncate to a single state per site.
Begin by dividing $H$ as follows:
 \begin{eqnarray}
        H &=& \sum_j \left(  H(j) + V(j) \right) ,\\
      H(j) &=& \frac{1}{2} \left( \Pi(j)^2 + 2 \phi(j)^2 \right) ,\\
      V(j) &=& - \phi(j) \phi(j+1).
   \label{freefielddecomp}
 \end{eqnarray}
Truncate by keeping only the ground-state
of $H(j)$ for each site $j$; {\it i.e.,} keeping the oscillator state
$\vert\omega(j)\rangle$
of frequency $\omega=\sqrt{2}$.   Note, this procedure truncates the
entire Hilbert space to a single product-state and therefore the
renormalized Hamiltonian will be a $1\times 1$-matrix, as will each
term in the expansion
 \begin{equation}
   H^{\rm ren} = \sum_{j,r} h_r(j)^{\rm conn} .
   \label{freeenergydensity}
 \end{equation}
Since the CORE procedure guarantees that $H^{\rm ren}$ has the same low
energy structure as the original theory, keeping only one state means
that we will only be able to compute the ground-state energy of the
free scalar-field theory. We will see that all of the terms
$h_r(j)^{\rm conn}$ are independent of $j$ and so it follows from
Eq.~\ref{freeenergydensity} that the ground-state energy density will
be given by
 \begin{equation}
   {\cal E}_{\rm free-field} = \sum_{r=1}^{\infty} h_r(j)^{\rm conn}
   \label{gsenergydensityform}
 \end{equation}
for any fixed $j$.

Following the basic rules, truncate $H(j)$ to obtain
 \begin{equation}
   h_1(j)^{\rm conn} = \frac{1}{2} \sqrt{2} ,
 \end{equation}
where $h_1(j)^{\rm conn}$ can be thought of as either a $1\times
1$-matrix or as a c-number.

To compute the range-2 contribution to the energy density, we must
diagonalize the two-site Hamiltonian
 \begin{equation}
   H(j)_{\rm 2-site} = \frac{1}{2} \left( \Pi(j)^2 + 2 \phi(j)^2
      + \Pi(j+1)^2 + 2 \phi(j+1)^2 \right)
      - \phi(j) \phi(j+1) ,
   \label{twositeham}
 \end{equation}
and expand the tensor product state
$\vert\omega(j)\rangle\otimes\vert\omega(j+1)\rangle$
in terms of the eigenstates
of $H(j)_{\rm 2-sites}$. Since this tensor-product state has the exact
two-site ground state appearing in its expansion in terms of the two-site
eigenstates, $H_{\cal T}^{(2)}(j)$ is a $1\times 1$ matrix whose
single entry is the exact ground-state energy of $H_{\rm 2-site}$;
{\it i.e.,} $E_2 = \frac{1}{2} ( \sqrt{3} + 1 )$.
Furthermore, since $S$ has
to be a $1\times 1$ orthogonal matrix it is trivial.  It follows
from these facts that the connected range-2 contribution to the
ground-state energy density is given by
 \begin{equation}
   h_2(j)^{\rm conn} = E_2(j) - 2 h_1(j)^{\rm conn}
       = \frac{1}{2} ( \sqrt{3} + 1 - 2\sqrt{2} ).
   \label{examplesffetwoconn}
 \end{equation}

To construct the range-3 term, find the ground-state energy of the
three-site problem, $E_3$, and then subtract twice the range-2
contribution, because we can embed a connected two-site sublattice in
the three-site lattice in two ways, and three times the range-1
contribution, because the single-site can be embedded in the three-site
sublattice in three ways; {\it i.e.,}
 \begin{equation}
   h_3^{\rm conn}(j) = E_3(j) -2 h_2(j)^{\rm conn}
      - 3 h_1(j)^{\rm conn}.
   \label{threesiteconn}
 \end{equation}
To compute the range-$r$ contributions,
find the exact ground-state energy of the $r$-site
Hamiltonian, $E_r$, and then subtract the lower order $s$-range connected
contributions as many times as the corresponding connected $s$-site
sublattice can be embedded in the $r$-site problem:
 \begin{equation}
   h_r(j)^{\rm conn} = E_r(j) - 2 h_{r-1}(j)^{\rm conn}
       -  3 h_{r-2}(j)^{\rm conn} - \ldots - r h_1(j)^{\rm conn} .
   \label{genconnected}
 \end{equation}

We wish to emphasize the unusual nature of this formula in that we
calculate the energy density of the infinite-volume Hamiltonian system
by exactly solving a series of finite-lattice problems, each defined
with open boundary conditions, and recombine these results to cancel out
finite-volume effects.  The results shown in Table~\ref{fftable} show
the way in which the partial sums
   \begin{equation}
      \epsilon_n = \sum_{r=1}^n h_r(j)^{\rm conn}
      \label{ffpartialenergy}
   \end{equation}
converge to the true ground-state energy density.  The surprising
result, given that the energies $E_r$ are computed for problems with
open boundary conditions, is that the finite-volume effects appear to
cancel to order ${\cal O}(1/r^3)$, rather than ${\cal O}(1/r)$ as one
would expect for a theory defined on a finite lattice with open
boundary conditions, or like ${\cal O}(1/r^2)$ which one would expect
for a theory defined with periodic boundary conditions.  At this time
we do not completely understand why the convergence is this rapid, but
this behavior is seen in all of the examples we have studied.

\subsection{Two-State Truncation: Heisenberg Antiferromagnet}

There are several reasons for studying the Heisenberg antiferromagnet.
First, the model exhibits spontaneous breaking of a continuous symmetry
in two and three spatial dimensions and, although in one spatial
dimension the Mermin-Wagner\cite{MerminWagner} theorem forbids a
nonvanishing order parameter, the theory still has a massless particle;
it is interesting to see if we can obtain the ground-state
energy density, the massless spectrum, and the vanishing of the
staggered magnetization by means of a simple CORE computation. Second,
this theory is exactly solvable by means of the Bethe
ansatz\cite{Betheansatz} and so we can compare our results to the
exact ground-state energy density $\epsilon_{\rm exact} = -\ln(2) +
1/4 = -0.443147$.  Third, there is a computation by Anderson, based
on an approximate spin-wave computation, which reproduces the
spin-1/2 antiferromagnet energy-density to within 2.5\%.  Although
this approximate result is based on treating the spin-1/2 system as
if it had spin-$N$, for $N \gg 1$, and then evaluating the result for
$N=1/2$, it has been difficult to do as well by earlier Hamiltonian
real-space renormalization group methods; we are finally able to
exhibit a simple approximate CORE computation which does significantly
better than Anderson's spin-wave computation working with spin-1/2 from
the outset.  The final reason for studying this case is that the
symmetry of the model makes it possible to describe the details of the
computation in a straightforward manner.  In particular, it is simple
to explain the need for, and construction of, the triangulation
transformation $S$ which we referred to when we stated the basic
rules for doing a CORE computation.

The Heisenberg antiferromagnet is a theory with a spin-1/2 degree of
freedom $\vec{s}(j)$ attached to each site $j$ of a one-dimensional
spatial lattice and a nearest-neighbor Hamiltonian of the form
 \begin{equation}
   H = \sum_j \, \vec{s}(j)\cdot\vec{s}(j+1).
   \label{HAFHam}
 \end{equation}
The $\vec{s}(j)$'s are operators which act in the single-site Hilbert
spaces ${\cal H}_j$ and satisfy the familiar angular momentum
commutation relations
 \begin{equation}
   \left[ s_{\alpha}(j), s_\beta(l) \right] = i \delta_{jl}
   \epsilon_{\alpha \beta \gamma } s_\gamma(j).
  \label{spinalg}
 \end{equation}

To analyze this problem, divide the lattice into three-site blocks
and label each block by an integer $j$.  The sites within each block
are labelled by the integers $\{3j,3j+1,3j+2\}$. Corresponding to this
decomposition of the lattice into blocks, divide the Hamiltonian into
two parts, $H_{\rm B}$ and $V_{\rm BB}$,
 \begin{eqnarray}
      H_{\rm B} &=& \sum_j H_{\rm B}(j) = \sum_j \,\left[\,
      \vec{s}(3j)\cdot\vec{s}(3j+1) +
            \vec{s}(3j+1)\cdot\vec{s}(3j+2) \, \right]  , \\
      V_{\rm BB} &=& \sum_j V_{\rm BB}(j) = \sum_j
      \vec{s}(3j+2)\cdot\vec{s}(3j+3).
   \label{HAFdecomp}
 \end{eqnarray}
Truncate by keeping the two lowest-lying
eigenstates of $H_{\rm B}(j)$ for each block ${\rm B}(j)$ so as to
produce a new coarser lattice which again has a spin-1/2 degree of
freedom associated with each of its sites. Diagonalizing $H_{\rm B}(j)$
is a simple exercise in coupling three spins; {\it i.e.,}
 \begin{eqnarray}
   H_{\rm B}(j) &=& \vec{s}(3j)\cdot\vec{s}(3j+1) +
    \vec{s}(3j+1)\cdot\vec{s}(3j+2), \\
      &=& \vec{s}(3j+1)\cdot\left(\vec{s}(3j)+\vec{s}(3j+2) \right), \\
      &=& \frac{1}{2}\left(S_{\rm tot}^2(j) - S_{(0+2)}^2(j)
          -3/4\right),
   \label{HAFthrees}
 \end{eqnarray}
where $\vec{S}_{\rm tot}(j) = \vec{s}(3j)+\vec{s}(3j+1)+\vec{s}(3j+2)$ and
$\vec{S}_{0+2}(j)= \vec{s}(3j) + \vec{s}(3j+2)$.  From Eq.~\ref{HAFthrees}
we see that the eigenstates of $H_{\rm B}(j)$ can be labelled by
the eigenvalues of
$S_{\rm tot}^2(j)$ and $S_{(0+2)}^2(j)$, and the two lowest-lying
eigenstates belong to the spin-1/2 multiplet for which the spins on sites
$3j$ and $3j+2$ couple to spin-1.  We denote these two degenerate
states by $\vert\uparrow_j\rangle$ and $\vert\downarrow_j\rangle$
and use them to construct the projection operator
 \begin{equation}
  P = \prod_j P(j); \quad
         P(j) = \vert\uparrow_j\rangle\langle\uparrow_j\vert
         + \vert\downarrow_j\rangle\langle\downarrow_j\vert.
   \label{HAFssprojops}
 \end{equation}
Using the $P(j)$'s we construct the connected range-1 operators
 \begin{equation}
   h_1(j)^{\rm conn} = P(j) H_{\rm B}(j) P(j) = - {\bf 1}(j),
   \label{hiconn}
 \end{equation}
where ${\bf 1}(j)$ stands for the $2\times 2$ identity matrix.

To obtain the connected range-$2$ term $h_2(j)^{\rm conn}$,
construct the Hamiltonian for the two-block or six-site problem. Since
this Hamiltonian commutes with the total-spin operators for the
six-site sublattice, the eigenstates of $H_{\rm six-sites}$ will fall
into spin-3, spin-2, spin-1 or spin-0 multiplets.
The following state,
\begin{equation}
    \frac{1}{\sqrt{2}} \left( \vert\uparrow_j\downarrow_{j+1}\rangle
         -\vert\downarrow_j\uparrow_{j+1}\rangle \right),
\end{equation}
is the unique linear combination of the original tensor-product states
which has total spin zero; hence, only spin-0 states appear in the
expansion of this state in terms of eigenstates of $H_{\rm six-sites}$.
The lowest-lying eigenstate of $H_{\rm six-sites}$ appearing in
the expansion of this spin-0 state is the ground-state of $H_{\rm
six-sites}$ whose eigenvalue we denote by $\epsilon_0$.  Similarly, the
following states
\begin{equation}
      \vert\uparrow_j\uparrow_{j+1}\rangle \quad , \quad
       \frac{1}{\sqrt{2}}\left( \vert\uparrow_j\downarrow_{j+1}\rangle
         + \vert\downarrow_j\uparrow_{j+1}\rangle \right) \quad , \quad
            \vert\downarrow_j\downarrow_{j+1}\rangle,
\end{equation}
are linear combinations of the original tensor-product states which
have total spin 1 and total $z$-component of spin $M_z=+1, 0, -1$,
respectively.  The lowest-lying eigenstate of $H_{\rm six-sites}$
appearing in each of these spin-1 combinations is that member of the
lowest-lying spin-1 multiplet having the appropriate value of $M_z$;
hence, each of these states contracts onto a unique eigenstate of
$H_{\rm six-sites}$. If we denote the degenerate eigenvalue of these
eigenstates by $\epsilon_1$, then the operator $H_{\cal T}^{(2)}(j)$
has the form
 \begin{equation}
 H_{\cal T}^{(2)}(j)= \left(
 \begin{array}{cccc}
    \epsilon_0 &   0 &  0  &  0 \\
     0  & \epsilon_1 &  0  &  0 \\
     0  &   0 & \epsilon_1 &  0 \\
     0  &   0 &   0 & \epsilon_1
  \end{array}
   \right),
   \label{hdmatrix}
 \end{equation}
using these remnant eigenstates as our new basis states. We could
use the explicit form of the triangulation matrix $S$, which rewrites
the original tensor product states in terms of these spin eigenstates, to
transform this back into the original tensor-product basis
 \begin{equation}
   \vert\uparrow_j \uparrow_{j+1}\rangle,
   \vert\uparrow_j \downarrow_{j+1}\rangle,
   \vert\downarrow_j \uparrow_{j+1}\rangle,
   \vert\downarrow_j \downarrow_{j+1}\rangle,
   \label{newspinstates}
 \end{equation}
but this is unnecessary since symmetry considerations require $S^{\dag}
H_{\cal T}^{(2)}(j) S$ to have the form
 \begin{equation}
   \beta_0 {\bf 1}_j \otimes {\bf 1}_{j+1}
      + \beta_1 \vec{s}(j) \cdot \vec{s}(j+1) .
   \label{newvalueofhd}
 \end{equation}
Eq.~\ref{newvalueofhd} can be rewritten
in terms of the total spin operator
for sites $j$ and $j+1$ to obtain
 \begin{equation}
   \epsilon_0 = \beta_0 - \frac{3}{4}\beta_1 \quad ; \quad
         \epsilon_1 = \beta_0 + \frac{1}{4} \beta_1 .
   \label{evals}
 \end{equation}
While the symmetry of this system makes it possible to determine
$\epsilon_0$ and $\epsilon_1$ analytically, it is more convenient to
compute it numerically.  To six significant figures, this calculation
yields
 \begin{equation}
   \epsilon_0 = -2.493577 \quad ; \quad \epsilon_1 = -2.001995.
   \label{epsilonvals}
 \end{equation}

To construct the connected range-2 term, we subtract the two ways of
embedding the one-block sublattices into the connected two-block
sublattice
 \begin{equation}
   h_2(j)^{\rm conn} =  (\beta_0 + 2) {\bf 1}_j \otimes {\bf 1}_{j+1}
      + \beta_1 \vec{s}(j) \cdot \vec{s}(j+1) .
   \label{twositeconnectedhaf}
 \end{equation}
We could go on to compute range-$r$ connected terms for $r > 2$,
but we will stop at range-2 and define the approximate renormalized
Hamiltonian by
 \begin{equation}
   H_{\rm ren} = \sum_j C {\bf 1}_j + \beta_1 \vec{s}(j) \cdot
      \vec{s}(j+1),
   \label{hafrenhamiltonian}
 \end{equation}
where $C = \beta_0+1$.  Clearly this approximate Hamiltonian, except
for the trivial addition of a multiple of the unit matrix, has the same
form as the original Hamiltonian.  When this happens, we say that the
theory is at a critical point, and $\vert\beta_1\vert < 1$ implies that
it has no mass gap. ( The logic which says that $\vert\beta_1\vert < 1$
implies no mass gap is that if we iterate the renormalization group
transformation, then eventually only the c-number part of the
Hamiltonian will remain. Since the interaction part becomes vanishingly
small, eventually all of the low-energy states of the theory must have
a vanishingly small energy splitting. Hence, the theory must have a
vanishingly small mass gap.)

To extract the ground-state energy density, we have to pay attention to
the constant term.  After the first transformation, we see that this
term will make a contribution to the ground-state energy density equal
to $C/3$, where the factor of $1/3$ appears because each site on the
new lattice corresponds to three sites of the old lattice.  Remembering
this and performing the renormalization group transformation on
the interaction term $\beta_1 \vec{s}(j) \cdot \vec{s}(j+1)$,
we generate a new renormalized Hamiltonian of the form
 \begin{equation}
   H'_{\rm ren} = \sum_j \beta_1 C {\bf 1}(j) + \beta_1^2
            \vec{s}(j) \cdot \vec{s}(j+1).
   \label{secondrenham}
 \end{equation}
Accumulating the new constant $\beta_1\ C/9$ into the previous
computation of the energy-density (where the $1/9$ comes from the fact
that one point on the new lattice corresponds to nine points on the
original lattice), we again have a new Hamiltonian which has the same
form as the original Hamiltonian, except that it is multiplied
by the factor $\beta_1^2$.  Repeating this process an infinite number
of times yields a series for the ground-state energy density
 \begin{equation}
   {\cal E}_{\rm ren-grp} =  \frac{C}{3} \sum_{n=0}^{\infty} \left(
   \frac{\beta_1}{3} \right)^n = \frac{C}{3(1-\beta_1/3)} = -0.4484462,
  \label{gsenergydensity}
 \end{equation}
which agrees well with the exact result ${\cal E}_{\rm exact} =
-0.443147$.   Thus, this simple range-2 calculation gives a result
which is good to about one-percent; this is more than a factor of two
better than that obtained from Anderson's spin-wave computation. Note
that this very simple calculation yields the {\em exact} mass gap.  One
also finds that the staggered magnetization ${\cal M} = \sum_j (-1)^j
s_z(j)$ vanishes (note that one obtains a nonvanishing staggered
magnetization on two- and three-dimensional spatial lattices).

This completes our present discussion of the antiferromagnet.  We will
return to it again in the section on questions of convergence since it
has something to teach us about the reliability of single-state
truncation calculations.

\subsection{Two-State Truncation: The $1+1$--dimensional Ising Model}

We now revisit the $1+1$-dimensional Ising model which we discussed in
Ref.~\cite{COREPrl} using an earlier formulation of the
CORE approximation.  While our earlier treatment was quite successful
in extracting the physics of the model, our new approach produces better
results, is less computationally intensive, and is much easier to
implement and explain.  There are two main reasons for treating this
example in some detail.  First, remarkably accurate results can be
obtained even when considering only terms up to range-3 in the
renormalized Hamiltonian.  Secondly, this problem does not have the
high degree of symmetry of the Heisenberg antiferromagnet and so
the construction of the operator $S$ must be done explicitly.

The Hamiltonian of the 1+1-dimensional Ising model is
 \begin{eqnarray}
   H_{\rm Ising} &=& -\sum_j\,\left[ c_\lambda \sigma_z(j) + s_\lambda
              \sigma_x(j)\sigma_x(j+1) \right] ,  \label{ISham}\\
         c_\lambda &=& \cos(\lambda \pi/2),  \quad
         s_\lambda = \sin(\lambda \pi/2),\nonumber
   \end{eqnarray}
where $j$ labels the sites on the infinite one-dimensional spatial
lattice and $0 \le \lambda \le 1$.  This model is interesting for
several reasons.  First, it exhibits a second-order phase transition
at $\lambda=1/2$;
for $\lambda < 1/2$, the ground state of the system is unique, the
order parameter $\langle \sigma_x(j)\rangle$ vanishes and the excited
states are localized spin excitations; when $\lambda > 1/2$, the
ground state is twofold-degenerate corresponding to values of the order
parameter
$\langle\sigma_x(j)\rangle=\pm(1-\cot^2(\lambda\pi/2))^{1/8}$ and the
excitations are solitons (kinks and antikinks).
Secondly, the model is exactly
solvable and so we have exact results with which to compare.
Thirdly, the model has
much less symmetry than the Heisenberg model and so the structure of
the renormalization group transformation is richer.

In order to show how a more complicated approximate renormalization
group transformation works, we once again adopt a two-state, three-site
block truncation algorithm, but now we compute $h_r(j)^{\rm
conn}$ for $r=1, 2, 3$.  Because the Hamiltonian is more
complicated than that of the antiferromagnet and computing the
connected range-3 terms involves solving for the eigenstates of a
nine-site problem, we must resort to numerical methods to carry
out the computation.  We numerically diagonalize the $512 \times 512$
nine-site Hamiltonian matrix.   However, since we only need a few
low-lying states to compute $S$, we could significantly reduce the
computational cost by using the Lanczos method. While unnecessary
for this simple problem, the application of the
Lanczos method to the construction of $S$ will be very useful when
studying more complicated theories.

The starting Hamiltonian is invariant under parity and
the simultaneous transformation $s_x(j) \rightarrow -s_x(j)$.  Our
thinning algorithm preserves this symmetry so that
the most general form the {\em renormalized Hamiltonian} can take is
 \begin{eqnarray}
      H^{\rm ren} &=& - \sum_{\alpha,j} c_\alpha {\cal O}_\alpha(j), \\
      {\cal O}_\alpha(j) &=&
         \sigma_{\alpha_0}(j)\,\sigma_{\alpha_1}(j+1)\cdots
             \sigma_{\alpha_r}(j+r) ,
   \label{ISgenham}
 \end{eqnarray}
where the $c_\alpha$'s are the couplings, $\alpha$ labels the different
types of operators which can appear, and $j$ is a site label.  Given
the symmetries of the original Hamiltonian which will be preserved in
the renormalized Hamiltonian, we see that the only two possible
one-site operators are $\alpha^{(1)}=\lbrace u,z\rbrace$, where $u$
denotes the identity operator; in other words, the only one-site
operators are $O_u(i)=\sigma_u(i)={\bf 1}$ and $O_z(i)=\sigma_z(i)$.
Similarly, the only two-site operators which are consistent with
the symmetries of the problem are $\alpha^{(2)}=\lbrace xx,yy,
zz\rbrace$, and the only three-site operators which can appear are
$\alpha^{(3)}=\lbrace xzx,xux,xxz,zxx,yzy,yuy,yyz,$
$zyy,zuz,zzz\rbrace$.

Since the original form of the Hamiltonian given in
Eq.~\ref{ISham} is just
a special form of Eq.~\ref{ISgenham}, we will discuss the truncation
procedure for the general case.  Once again we work with blocks ${\rm
B}_j$ containing the points $\{3j,3j+1,3j+2\}$ and keep the lowest two
eigenstates of the generic block Hamiltonian
 \begin{eqnarray}
    H_B(j) &=& - c_z \bigl[ {\cal O}_{z}(3j) + {\cal O}_{z}(3j+1) +
      {\cal O}_z(3j+2)\bigr] - c_{xx} \bigl[ {\cal O}_{xx}(3j)
      + {\cal O}_{xx}(3j+1)\bigr] \nonumber   \\
      &-& c_{yy} \bigl[ {\cal O}_{yy}(3j) + {\cal O}_{yy}(3j+1)\bigr]
      - c_{zz} \bigl[ {\cal O}_{zz}(3j) + {\cal O}_{zz}(3j+1)\bigr]
      - c_{xux} {\cal O}_{xux}(3j) \nonumber\\
      &-& c_{yuy} {\cal O}_{yuy}(3j) - c_{zuz} {\cal O}_{zuz}(3j)
      - c_{xzx} {\cal O}_{xzx}(3j) - c_{xxz} {\cal O}_{xxz}(3j)
      - c_{zxx} {\cal O}_{zxx}(3j) \nonumber\\
      &-& c_{yzy} {\cal O}_{yzy}(3j) - c_{yyz} {\cal O}_{yyz}(3j)
      - c_{zyy} {\cal O}_{zyy}(3j) - c_{zzz} {\cal O}_{zzz}(3j).
   \label{ISHB}
  \end{eqnarray}
When we truncate $H_{\rm B}(j)$ to these two states, we obtain the new
range-1 terms.  Since $h_1(j)^{\rm conn}$ is a diagonal matrix in
this basis, it can be written as a sum of a multiple of the unit matrix
${\bf 1}(j)$ and a multiple of $s_z(j)$.

If we denote the eigenstates of the three-site block ${\rm B}_j$ by
$\vert\uparrow_j\rangle$ and $\vert\downarrow_j\rangle$,
then the connected range-2 contributions to the
renormalized Hamiltonian are obtained by first solving the two-block or
connected six-site problem exactly and expanding the four
tensor-product states
 $\vert\phi_\alpha({\rm B}_j,{\rm B}_{j+1})\rangle=\{$
 $\vert\uparrow_j\uparrow_{j+1}\rangle$,
 $\vert\uparrow_j\downarrow_{j+1}\rangle$,
 $\vert\downarrow_j\uparrow_{j+1}\rangle$ and
 $\vert\downarrow_j\downarrow_{j+1}\rangle\}$ in
terms of the eigenstates $\vert\epsilon_\beta({\rm B}_j,
{\rm B}_{j+1})\rangle$ of this problem.  To compute $S$ for the
range-2 problem, begin by constructing the $4 \times 64$ overlap matrix
$ Q_{\alpha\beta} = \langle\phi_\alpha({\rm B}_j,{\rm B}_{j+1})\vert
\epsilon_\beta({\rm B}_j,{\rm B}_{j+1})\rangle$.
Note that each row of $Q$ gives the expansion of each of the
four tensor-product states in terms of the eigenstates of $H_{\rm
six-site}$.  Ensure that the eigenstates are arranged in order of
increasing eigenenergy.

The construction of $S$ now proceeds iteratively.  Begin by
focusing attention on the first column of $Q$; this is a
$4\times 1$ matrix $C_1$ whose entries contain the overlaps
of the four tensor-product states with the eigenstate of lowest
energy.   If $C_1$ has any nonzero entries, then
we can find a rotation matrix $R_1$ such that $C_1$ can be brought
into a form where only its upper entry is nonzero.  Finding
such an $R_1$ is equivalent to constructing the singular value
decomposition of $C_1$.  Using $R_1$, transform $Q$ to
${\tilde Q}_1 = R_1\ Q$ and then focus attention on the $3\times
63$ submatrix obtained by eliminating the first row and first column
of ${\tilde Q}_1$; call the resulting matrix $Q_1$.

Apply the same reasoning to $Q_1$.  Focus on the first column of $Q_1$,
denoted by $C_2$.  If $C_2$ contains some nonvanishing entries,
construct the orthogonal $3\times 3$ transformation $U^\dagger_2$
which brings $C_2$ into the standard form where only the upper element
is nonvanishing.  Again, this is equivalent to performing the
singular value decomposition of $C_2$.  Next, define a $4\times 4$
matrix $R_2$ as
 \begin{equation}
   R_2 = \left(
  \begin{array}{cccc}
     1 & 0 & 0 & 0 \\
     0 & \cdots  &\cdots   & \cdots   \\
     0 & \vdots  & U^\dagger_2 &  \\
     0 & \vdots  &     &
     \end{array}
    \right).
   \label{rtwodef}
 \end{equation}
Transform $Q$ to ${\tilde Q}_2 = R_2 R_1 Q $, then construct the
$2\times 62$ submatrix $Q_2$ formed by eliminating the first two rows
and columns from ${\tilde Q}_2$.

Next, construct the $2\times 2$ matrix $U^\dagger_3$ which brings the
first column of $Q_2$ into standard form, extend $U^\dagger_3$ to a
$4\times 4$ matrix
 \begin{equation}
   R_3 = \left(
   \begin{array}{cccc}
     1 & 0 & 0 & 0 \\
     0 & 1 & 0 & 0 \\
     0 & 0  &\cdots  & \cdots  \\
     0 & 0  &\vdots  & U^\dagger_3
   \end{array}
   \right),
   \label{rthreedef}
 \end{equation}
and then define the triangulation matrix
$S^{(2)} = R_3 R_2 R_1$.  Also, define the diagonal
$4\times 4$ Hamiltonian $H_{\cal T}^{(2)}$ from the four lowest
eigenvalues of $H_{\rm six-sites}$.  Then the $4\times 4$ connected
range-2 operator in the renormalized Hamiltonian is given by
 \begin{equation}
   h_2(j)^{\rm conn} = S^{\dag(2)} H_{\cal T}^{(2)} S^{(2)} -
         h_1(j)^{\rm conn} - h_1(j+1)^{\rm conn} .
   \label{htwoconnising}
 \end{equation}

Note that we have simplified this discussion by assuming, as is
usually the case, that the eigenvalues of $H_{\rm six-sites}$ are
nondegenerate and that the four tensor-product states have
nonvanishing overlaps with the four lowest-lying eigenstates.  If this
is not the case, then we have to generalize this discussion slightly.
In the event that some eigenstates do not occur in the expansion of the
tensor product states, the corresponding matrix $Q_1$ or $Q_2$ will have
a first column in which all entries are zero.  When this happens, simply
eliminate this column and use the first nonvanishing column to define
the rotation matrix; the corresponding eigenvalue is then used in
$H_{\cal T}^{(2)}$.  When an eigenvalue is $g$-fold degenerate, include
in $C_1$ or $C_2$ all $g$ columns of $Q_1$ or $Q_2$ corresponding to
the eigenvectors in the degeneracy subspace and then carry out the
singular value decomposition $C_j=U_j \Sigma_j V^\dagger_j$.  The
required rotation matrix is again obtained from $U^\dagger_j$, but now
$V_j$ is needed to construct the remnant eigenstates from the
degeneracy subspace.  Taking degeneracies into account can become
important after a large number of iterations as the renormalized
Hamiltonian flows closer and closer to one of its fixed points.

Constructing the range-3 connected terms proceeds the same way, except
now we have to work with three adjacent blocks ${\rm B}_j$, ${\rm
B}_{j+1}$ and ${\rm B}_{j+2}$. Now the matrix $Q$ has eight columns
corresponding to the tensor-product states
$\vert\!\uparrow_j\uparrow_{j+1}\uparrow_{j+2}\rangle$,
$\vert\!\downarrow_j\uparrow_{j+1}\uparrow_{j+2}\rangle$,
$\vert\!\uparrow_j\downarrow_{j+1}\uparrow_{j+2}\rangle$,
$\vert\!\uparrow_j\uparrow_{j+1}\downarrow_{j+2}\rangle$,
$\vert\!\downarrow_j\downarrow_{j+1}\uparrow_{j+2}\rangle$,
$\vert\!\downarrow_j\uparrow_{j+1}\downarrow_{j+2}\rangle$,
$\vert\!\uparrow_j\downarrow_{j+1}\downarrow_{j+2}\rangle$, and
$\vert\!\downarrow_j\downarrow_{j+1}\downarrow_{j+2}\rangle$,
and we must compute (assuming nondegeneracy of the
spectrum and no missed states) the eight lowest eigenstates of $H_{\rm
nine-sites}(j)$ in order to construct $H_{\cal T}^{(3)}$ and $S^{(3)}$.
Except that we are dealing with slightly larger matrices, we go through
the same steps described above.  The range-3 contribution to
the renormalized Hamiltonian is then given by
 \begin{eqnarray}
      h_3(j)^{\rm conn} &=& S^{\dag(3)} H_{\cal T}^{(3)}(j) S^{(3)} -
         h_1(j)^{\rm conn} - h_1(j+1)^{\rm conn}- h_1(j+2)^{\rm conn}
       \nonumber \\
       &-& h_2(j)^{\rm conn} - h_2(j+1)^{\rm conn}.
   \label{hthreeconnising}
 \end{eqnarray}
Given $h^{(1)}(j)$, $h^{(2)}(j)$, and $h^{(3)}(j)$, the approximate
renormalized Hamiltonian on the lattice with one-third as many
sites as the original is
 \begin{equation}
   H^{\rm ren} = \sum_j \left[h_1(j) + h_2(j) + h_3(j)\right].
   \label{ISHren}
 \end{equation}

Our results were obtained by choosing a specific value of $\lambda$
in the special form of the Hamiltonian given by
Eq.~\ref{ISham} in which only $c^{(1)}_{z}$ and $c^{(1)}_{xx}$ differ
from zero.  We then apply the above range-3 CORE procedure.  After
the first RG transformation, we obtain a new Hamiltonian comprised
of all allowed operators with nonvanishing couplings; however, most of
the couplings are small.  We iterate the process, obtaining a
sequence of renormalized Hamiltonians in which the couplings flow
until finally, all but one of the coefficients vanish; {\it i.e.,}
until one reaches a solvable fixed-point Hamiltonian.   Our numerical
computations show that there are only two possible fixed-point
Hamiltonians: one in which only $c^{(\infty)}_{z}$ is nonvanishing,
and one in which only $c^{(\infty)}_{xx}$ is different from zero.

\begin{figure}
\begin{center}
\leavevmode
\epsfxsize=5in\epsfbox[80 160 530 760]{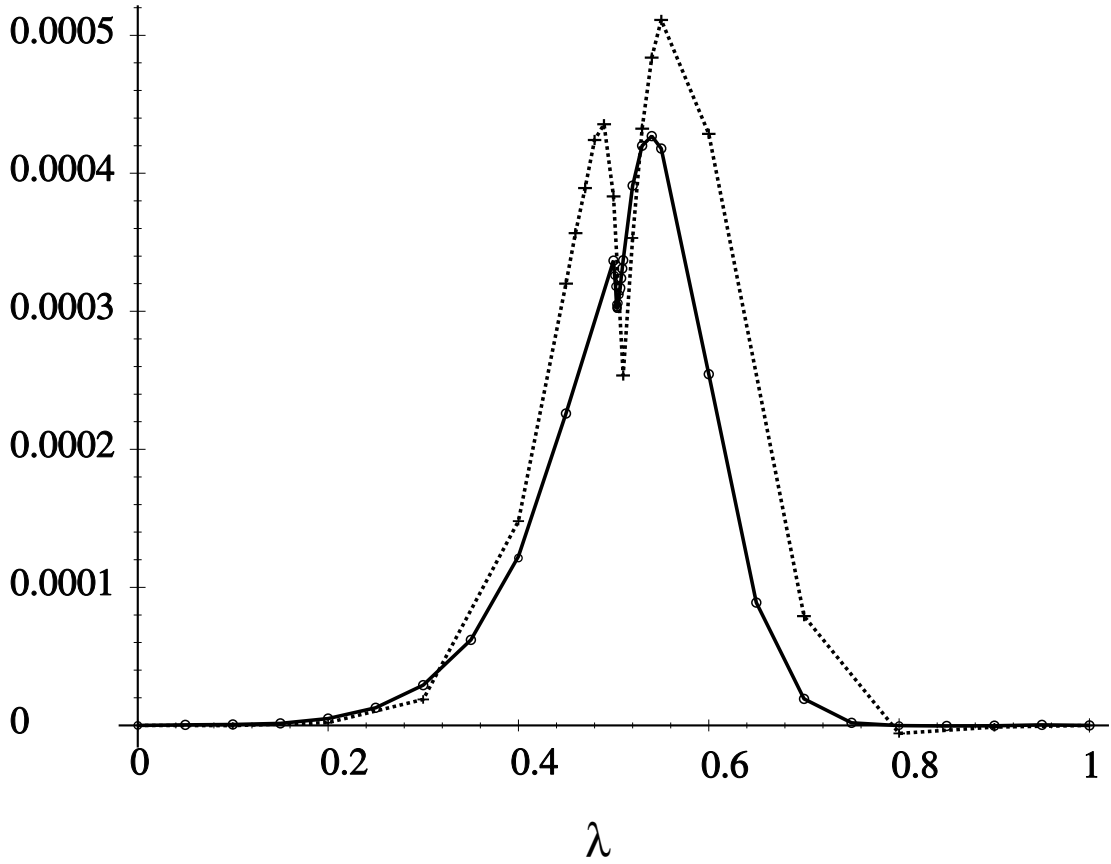}
\end{center}
\caption[ediff]{Fractional error in CORE estimates of the
ground-state energy density in the Ising model against
$\lambda$.  The dotted curve with crosses shows previous estimates
from Ref.~\protect\cite{COREPrl}; results from the present work are
shown by the solid curve with circles.}
\label{ISEDENS}
\end{figure}

In Fig.~\ref{ISEDENS}, we plot the fractional error in the CORE
estimates of the ground-state energy density. The dotted curve shows
the results obtained in Ref.~\cite{COREPrl} using the earlier version
of the CORE approximation.  The critical value $\lambda_c$ separating
the spontaneously-broken phase from the unbroken phase is found to be
$\lambda_c \approx 0.50365$ which agrees well with the exact value of
$1/2$.

\begin{figure}
\begin{center}
\leavevmode
\epsfxsize=5in\epsfbox[80 100 530 760]{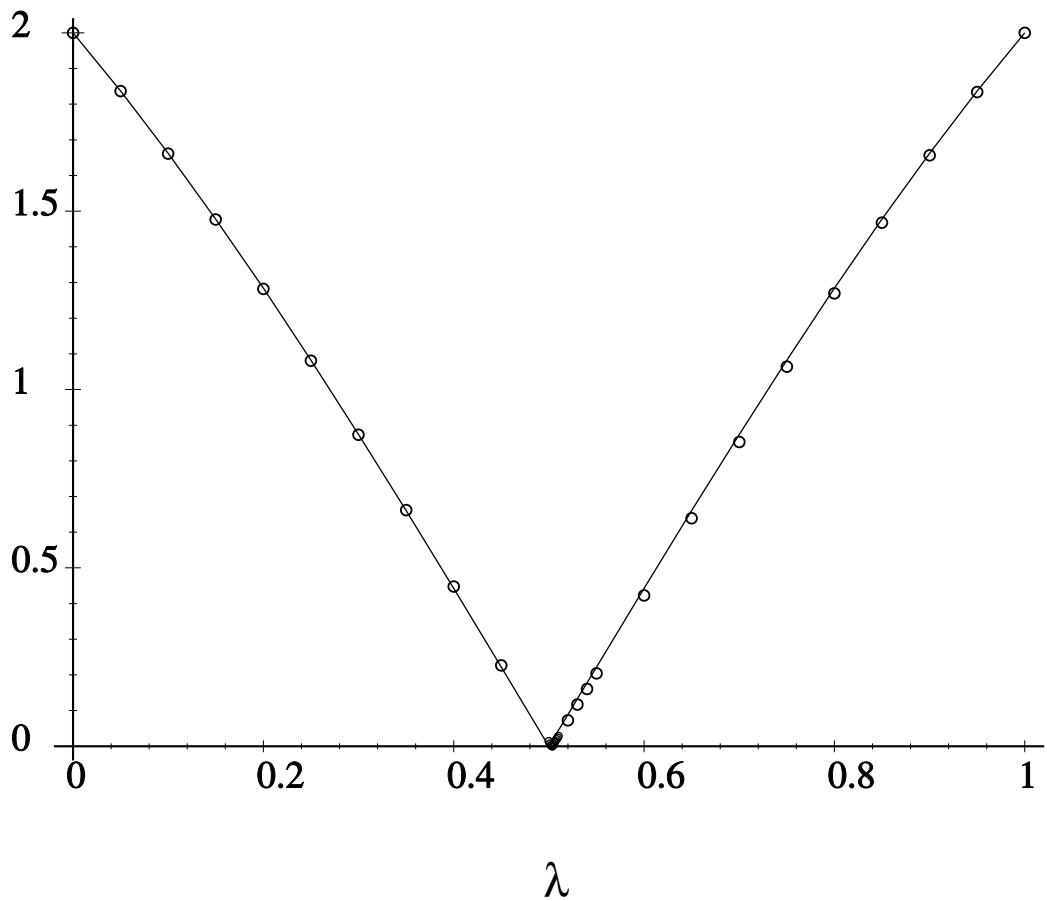}
\end{center}
\caption[massgap]{CORE estimates (circles) of the mass gap in the Ising
model against $\lambda$.  The solid curve shows the exact mass gap.}
\label{ISMASSGAP}
\end{figure}

Extracting the mass gap as a function of $\lambda$ is easily done
since both fixed-point Hamiltonians are exactly solvable.  Below the
phase transition where $c^{(\infty)}_{z}$ is the only nonvanishing
coefficient, eigenstates of the Hamiltonian are tensor products of
eigenstates of $\sigma_z(j)$, and so the mass gap is equal to $2
c^{(\infty)}_{z}$.  Above the phase transition, the only nonvanishing
coefficient is $c^{(\infty)}_{xx}$ and so eigenstates of the
Hamiltonian are products of eigenstates of $\sigma_x(j)$.  In this case,
there are two degenerate ground states; the discrete symmetry
$\sigma_x(j) \rightarrow -\sigma_x(j)$ is spontaneously broken.
In this phase, the low-lying eigenstates are {\em kinks} which have
mass $2 c^{(\infty)}_{xx}$.  The results of the CORE computations
for the mass gap are shown in Fig.~\ref{ISMASSGAP}.

\begin{figure}
\begin{center}
\leavevmode
\epsfxsize=5in\epsfbox[80 100 530 760]{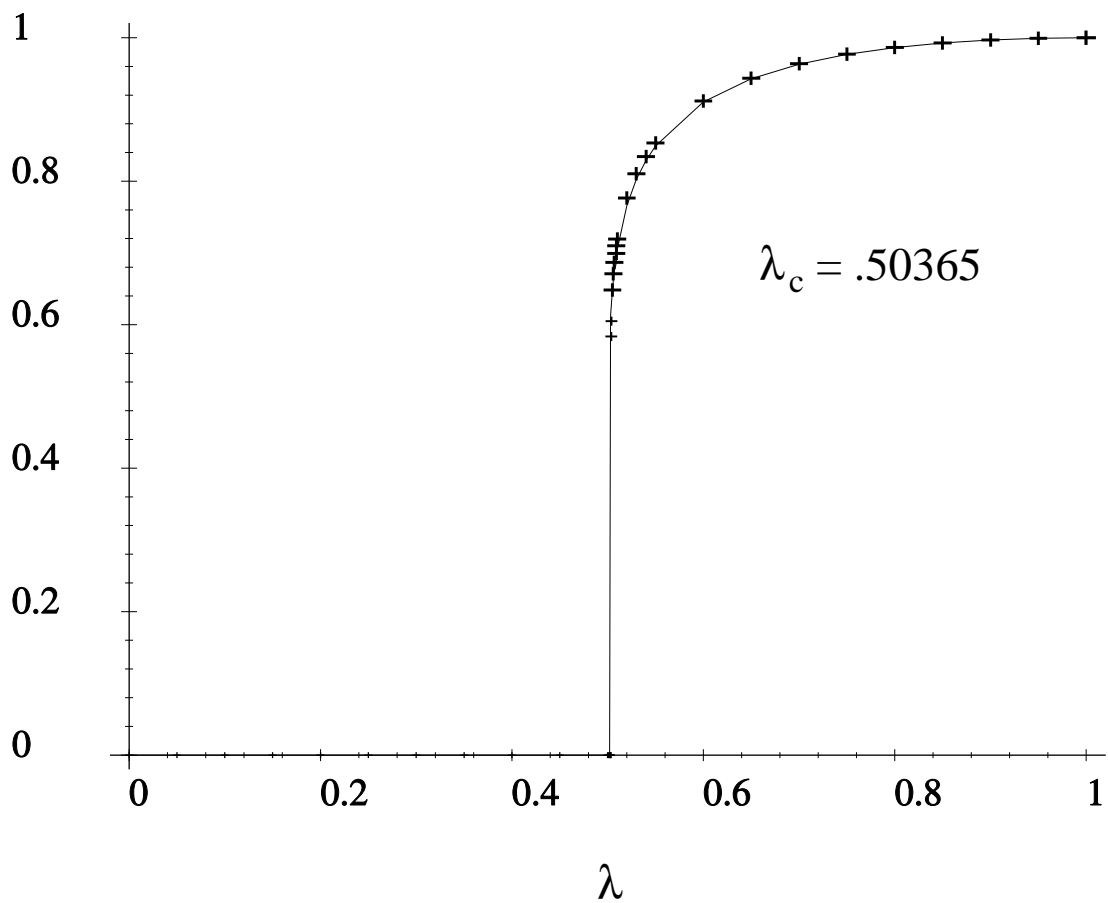}
\end{center}
\caption[magnet]{Comparison of the CORE estimates (crosses)
of the magnetization with the exact results (solid curve)
in the Ising model against $\lambda$.  $\lambda_c$ is the
critical point.}
\label{ISMAG}
\end{figure}

Finally, the magnetization was also studied.  A sequence of
renormalized magnetization operators was computed along with the
renormalized Hamiltonian; the starting operator in this CORE
sequence was $M = \sum_j \sigma_x(j)$.  The renormalized
magnetization $M^{\rm ren}$ has a cluster expansion given by
 \begin{equation}
   M^{\rm ren} = \sum_j \sum_{r=1}^{\infty} m_r(j)^{\rm conn},
   \label{ISMren}
 \end{equation}
where the connected range-$r$ operators $m_r(j)^{\rm conn}$ are
computed from the truncated one, two, and three-block operators,
 \begin{eqnarray}
      m_1(j)^{\rm conn} &=& M^{(1)}_{\cal T}(j), \\
      m_2(j)^{\rm conn} &=& S^{\dag(2)} M^{(2)}_{\cal T}(j) S^{(2)}
         - m_1(j)^{\rm conn} - m_1(j+1)^{\rm conn}, \\
      m_3(j)^{\rm conn} &=& S^{\dag(3)} M^{(3)}_{\cal T}(j) S^{(3)}
         - m_2(j)^{\rm conn} - m_2(j+1)^{\rm conn} \nonumber\\
               &-& m_1(j)^{\rm conn} - m_1(j+1)^{\rm conn}
               -m_1(j+2)^{\rm conn},
   \label{magforms}
 \end{eqnarray}
where $M^{(r)}_{\cal T}(j)$ consists of the matrix elements of
$M^{(r)}$ in the basis of remnant eigenstates, and $M^{(r)}$ is
the restriction of the full magnetization operator to the $r$-block
sublattice.  A comparison of the CORE estimates of the
magnetization with the exactly known results is shown in
Fig.~\ref{ISMAG}.

\begin{figure}
\begin{center}
\leavevmode
\epsfxsize=5in\epsfbox[80 160 530 760]{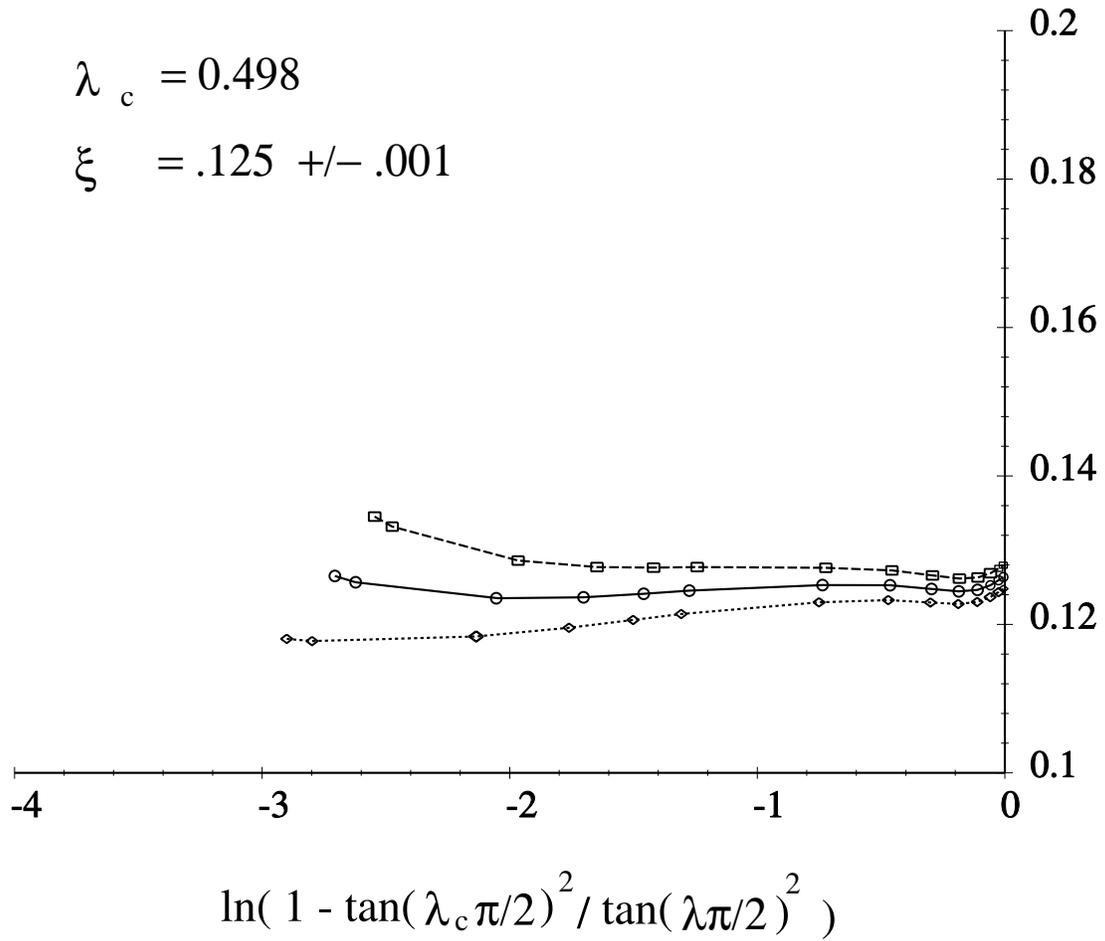}
\end{center}
\caption[exponent]{Plots of $p(\lambda)$ as given in
Eq.~\protect\ref{ISexpfit} for $\lambda_c=0.496$ (dashed curve
with squares), $0.498$ (solid curve with circles), and $0.500$
(dotted curve with diamonds).  $\lambda_c$ is the critical
coupling in the Ising model, and $\xi$ is the critical
exponent corresponding to the magnetization.}
\label{ISMAGTWO}
\end{figure}

Two procedures were used to extract the critical exponent from these
calculations.  Both procedures attempt to fit the logarithm of the
magnetization to the form of the exact answer, namely:
 \begin{equation}
   p(\lambda) = \frac{ \ln(M(\lambda)) }
   {\ln( 1 - \tan(\lambda_c\pi/2)^2
            / \tan(\lambda\pi/2)^2 )},
   \label{ISexpfit}
 \end{equation}
where $\ln(M(\lambda))$ stands for the logarithm of the computed values
of the magnetization and $\lambda_c$ stands for that value of $\lambda$
at which the theory changes phase.  If we attempt to extract
$p(\lambda)$ by fixing $\lambda_c = 0.50365$, the value above
which the CORE computation changes from having $c^{(\infty)}_{z} \ne 0$
to $c^{(\infty)}_{xx} \ne 0$, then the values of $p(\lambda)$ obtained
from this procedure do not lie on straight line.  Moreover, the average
value of $p(\lambda)$ lies between $0.10 - 0.11$, which is not a very
good fit to the exact value $1/8$.  If, on the other hand, we vary
$\lambda_c$ and determine its best value by fitting the resulting
values for $p(\lambda)$ to a straight line, then we obtain a very good
fit for $\lambda_c \approx 0.498$ and find that
$p(\lambda)$ lies in the range $0.1236 < p(\lambda) < 0.126 $.  The
discrepancy between $\lambda_c =  0.498$ and $\lambda_c =
0.50365$ gives {\it a priori} evidence, without knowledge of the exact
solution, that the determination of the critical point must have an
error of about one percent due to an accumulation of
numerical errors and limiting the computation to range-3 terms.
Fig.~\ref{ISMAGTWO} displays three plots of $p(\lambda)$ for $\lambda_c
= 0.496, 0.498,$ and  $0.500$;  the best fit to a straight line is
given by the middle curve which corresponds to $\lambda_c = 0.498$.

\subsection{Infinite-State Truncation: Free Scalar Field Theory}
\label{freebose}

Lastly, we return to the case of free scalar field theory
in $1+1$ dimensions, but this time, we use a truncation algorithm
which keeps an infinite number of states at each step.

Consider a truncation procedure based upon two-site blocks.
For each two-site block $B_p$, we introduce the operators
\begin{eqnarray}
     \Phi(p)_{\pm} &=& \frac{1}{\sqrt{2}} \left[ \phi(2p) \pm
        \phi(2p+1) \right], \\
     \Omega(p)_{\pm} &=& \frac{1}{\sqrt{2}} \left[ \Pi(2p) \pm
        \Pi(2p+1) \right],
\label{basicopsphifour}
\end{eqnarray}
and define ladder operators ${\cal A}_{+}(p)$ and
${\cal A}_{-}(p)$ using
\begin{eqnarray}
   \Phi(p)_{\pm} &=& \frac{1}{\sqrt{2\gamma_{\pm}}} \left[
                       {\cal A}(p)^{\dag}_{\pm} +
                       {\cal A}(p)_{\pm} \right], \\
   \Omega(p)_{\pm} &=& i \sqrt{\frac{\gamma_{\pm}}{2}} \left[
                       {\cal A}(p)^{\dag}_{\pm} -
                       {\cal A}(p)_{\pm} \right],
  \label{annihcreatphifour}
\end{eqnarray}
where $\gamma_{-} = \sqrt{\mu^2+ 3}$ and $\gamma_{+}=\sqrt{\mu^2+1}$.
In terms of these variables, the two-site Hamiltonian is simply
a sum of two decoupled oscillators, and its eigenstates are given by
   \begin{equation}
      \vert N_{+}(p),N_{-}(p)\rangle
 = \frac{1}{\sqrt{N_{+}!}\sqrt{N_{-}!}}\,\,
      {{\cal A}^{\dag}_{+}}(p)^{N_{+}} {{\cal A}^{\dag}_{-}}(p)^{N_{-}}
      \vert\gamma_{+},\gamma_{-}\rangle,
      \label{twositeeigens}
   \end{equation}
where ${\cal A}_{+}\vert\gamma_{+},\gamma_{-}\rangle
={\cal A}_{-}\vert\gamma_{+},\gamma_{-}\rangle=0$.
We now adopt a simple truncation procedure in which we keep an
infinite set of block-states
   \begin{equation}
      \vert N(p)\rangle = \vert N_{+}(p)=N;\ N_{-}(p)=0\rangle .
      \label{truncruleosc}
   \end{equation}
In other words, only states for which the higher-frequency
$\gamma_{-}$ oscillator is in its ground state are retained.
With this choice of eigenstates, $h_1(p)^{\rm conn}$
can be written as
   \begin{eqnarray}
      h_1(p)^{\rm conn} &=& \frac{\gamma_{-}}{2} +
       \gamma_{+}\left[ {\cal A}^\dagger_{+}(p) {\cal A}_{+}(p)
       + \frac{1}{2}\right],\\
       &=& \frac{\gamma_{-}}{2} + \frac{1}{2}\left[
       \Omega_{+}(p)^2 + \gamma_{+}^2\Phi_{+}(p)^2\right].
      \label{htwositeone}
   \end{eqnarray}

Now consider an $r$-block sublattice $B(p,\cdots,p+r-1)
=\{2p,2p+1,\cdots,2p+2r-1\}$.  The Hamiltonian restricted to
this sublattice has the form
   \begin{equation}
      H_r(p) = \frac{1}{2} \sum_{j=2p}^{2p+2r-1} \Pi(j)^2
      + \frac{1}{2} \sum_{j_1,j_2=2p}^{2p+2r-1} \phi(j_1)\ M_r(j_1,
      j_2)\ \phi(j_2),
      \label{quadformr}
   \end{equation}
where $M_r$ is a $2r\times 2r$ real-symmetric matrix whose
elements satisfy $M_r(i,j)={\cal M}_r(i-j)$.
$M_r$ can be diagonalized to obtain the normal modes, and
the $r$-lowest eigenvalues of $M_r$ then yield $H_{\cal T}$,
the remnant eigenvalues.  The triangulation matrix $S$ is
determined as usual, except that we can now work in terms of
the fields instead of basis states.  The proof of these statements
is a straightforward exercise in normal-ordering using simple
generalizations of the identities given in Appendix~\ref{app:sho}
and the definition of the operator $S$.  The connected range-$r$
term $h_r(p)^{\rm conn}$ is then computed from $H_r(p)$ by
subtracting from it the previously computed connected
range-$j$ terms, for $j = 1,\dots,r-1$.  Finally, the terms
$h_r(p)^{\rm conn}$ are combined to form the renormalized
Hamiltonian, which takes the form
\begin{equation}
H^{\rm ren} = \sum_p \frac{1}{2}\left[ h_0 + \Omega_+(p)^2
+ 2\sum_{m=0}^{r-1}
 \ \alpha_m^{(r)}\ \Phi_+(p)\Phi_+(p+m)
\right],
\end{equation}
where $h_0$ and $\alpha_m^{(r)}$ are $c$-numbers.
For example, if we truncate the cluster expansion after two-block
clusters ($r\!=\! 2)$, we find in each RG step $n\rightarrow n+1$
that $\alpha_1^{(2)}(n\!+\!1)=\alpha_1^{(2)}(n)/2$
and $\alpha_0^{(2)}(n\!+\!1)=\alpha_0^{(2)}(n)+(\sqrt{5}-1)
\alpha_1^{(2)}(n)/2$, for $\alpha_1^{(2)}(n) <0$.  Amazingly,
no $\Phi^4$ terms appear in the renormalized Hamiltonian, but
note that the exact CORE transformation of the nearest-neighbor
Hamiltonian results in a new Hamiltonian which has an infinite
number of terms.  The importance of these results is that
it shows we can, both in principle and in practice, directly deal
with field theories having an infinite number of states per site,
without first mapping them to spin systems.
A more detailed description of the above calculation will
appear in a forthcoming paper, which will also describe the
analogous calculation for Fermi fields.

\begin{figure}
\begin{center}
\leavevmode
\epsfxsize=5in\epsfbox[80 160 530 760]{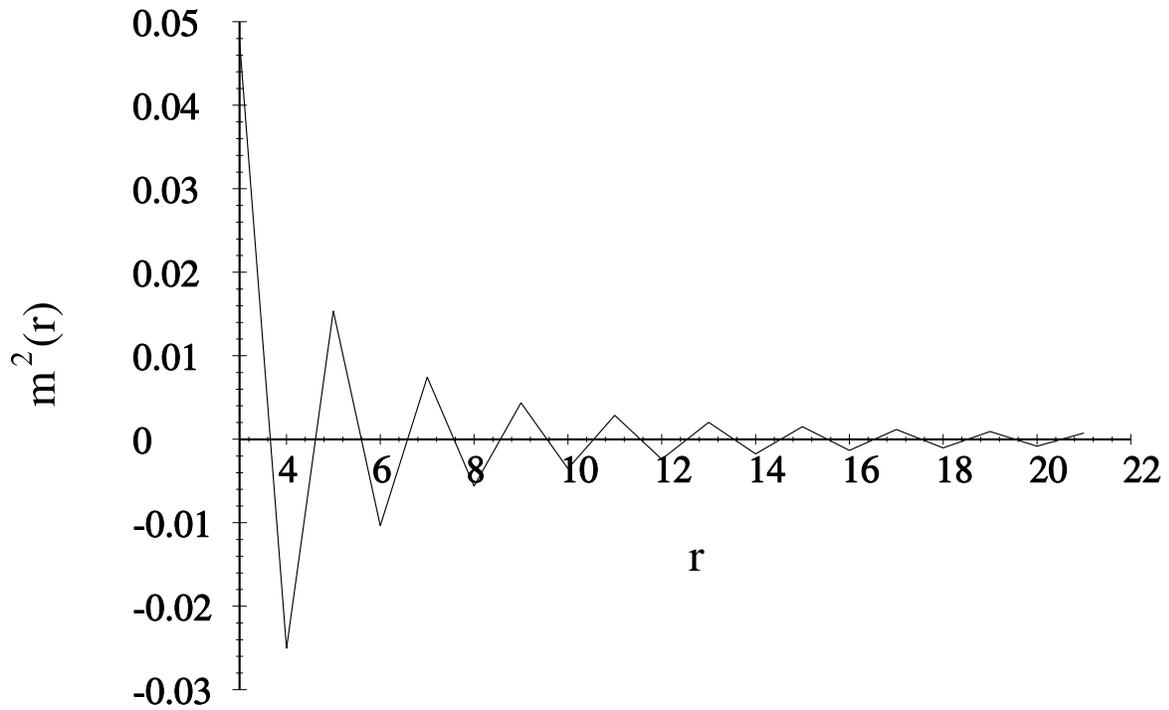}
\end{center}
\caption[ffmassgap]{CORE estimates of the mass gap squared $m^2(r)$
in the free scalar field theory against the truncation order $r$ in
the cluster expansion of the renormalized Hamiltonian.  The CORE
estimates are obtained using an infinite-state truncation algorithm.}
\label{FFMASSGAP}
\end{figure}

At this point, an interesting question is ``How big must $r$ be in
order to do a good job of reproducing the mass gap and correlation
functions of the free-field theory?''.  To analyze this question for
the massless field (the hardest case), expand the fields $\Phi_+$ and
$\Omega_+$ in terms of their Fourier components to rewrite the
renormalized Hamiltonian for $\mu=0$ as
 \begin{equation}
    H^{\rm ren} = \sum_{k=-\pi/2}^{\pi/2} \frac{1}{2}
     \left( \tilde\Omega(k) \tilde\Omega(-k) +
       2\left[\sum_{s=0}^{r-1} \alpha_s^{(r)} \cos(sk)\right]
       \ \tilde\Phi(k) \tilde\Phi(-k) \right).
    \label{hrfourier}
 \end{equation}
We then explicitly compute the $\alpha_s^{(r)}$ couplings for various
values of $r$, the cluster expansion truncation order.
Table \ref{alphatable} compares the values of $\alpha_0^{(r)}$,
$\alpha_1^{(r)}$, $\alpha_2^{(r)}$, and $\alpha_3^{(r)}$
obtained from a range-2, range-3,
and range-4 CORE computation.  We see from the table that any given
coefficient converges rapidly in $r$ to its $r\rightarrow \infty$
limit.  In this limit, the renormalized Hamiltonian matches
the original theory restricted to the subspace spanned by the
oscillators having momenta $-\pi/2 \le k \le \pi/2$; hence, we
can use
 \begin{equation}
   \sum_{s=0}^{\infty} \alpha_s^{(\infty)} \cos(sk) = 1-\cos(k/2),
    \label{matchingcond}
 \end{equation}
to determine the couplings in the $r\rightarrow \infty$ limit.
We find
 \begin{eqnarray}
   \alpha_0^{(\infty)} &=&  1-\frac{2}{\pi},  \\
   \alpha_s^{(\infty)} &=& \frac{4(-1)^s}{\pi(4s^2-1)}. \quad (s>0)
   \label{alphaequations}
 \end{eqnarray}
Note that the exact coefficients $\alpha_s^{(\infty)}$ fall
off as $1/s^2$.  This means that if we truncate the formula for the
frequency
 \begin{equation}
   \omega^2_r(k) = 2 \sum_{s=0}^{r-1} \alpha_s^{(\infty)} \cos(ks),
   \label{truncenergy}
 \end{equation}
then the mass $m(r)$, defined by $m^2(r)=\omega^2_r(0)$,
fails to vanish; in
fact, for some values of $r$, it becomes negative; the behavior of the
gap as a function of $r$ is shown in Fig.\ref{FFMASSGAP}.
Since negative values for the gap make no sense as they imply that
the renormalized Hamiltonian has no ground state, we can only
truncate after an even number of terms.  An important
observation we can make from the plot is that we can only
accurately compute the mass gap for the free field theory if we work
out to a range $r \approx 1/m $.  We should note, however, that
the mass is the quantity which is most sensitive to making a
finite-range truncation of the exact renormalized Hamiltonian and that
the function $\omega_r(k)$ converges for $k > m$ more rapidly
than it does for $k=0$.  This means that {\em even for the worst case}
of a free-field theory, careful examination of the behavior of the
function $\omega_r(k)$ and the oscillatory behavior of $m(r)$ will
allow us extract the correct physics without having to compute
an infinite number of terms in the finite-range cluster expansion.
A final point which we state without proof is that the importance
of the longer range terms is significantly reduced if we work with
larger blocks.  In one dimension for blocks of size $L$, the
corresponding coefficients $\alpha_s(r)$ fall off as $1/r^2L^2$.

This discussion applies to the truncation procedure in which we keep
an infinite set of states at each truncation step.  It would be
interesting to discover how much of this behavior occurs if we
keep only a finite number of states at the first step and map the
theory into a spin system.  The results of the Ising calculation
show that we do not always need to work with large values of $r$
in order to correctly reproduce the mass gap for a
theory near its critical point.

Note that CORE's ability to reproduce the mass gap and density of
states near zero momentum is much greater than that of the naive
$(t\!=\!0)$ renormalization group procedure. For example, if we
diagonalize the single-site Hamiltonian of the massless free scalar
field theory and keep the single-site ground state and first excited
state, then the naively-determined renormalized Hamiltonian is the
simple Ising Hamiltonian for a value of the coupling far from its
critical point;  the system has a nonvanishing mass gap.  However, if
we keep the same single-site states but use the CORE rules to construct
the renormalized Hamiltonian, many more interaction terms emerge and
the couplings in the renormalized Hamiltonian are such that the system
is much closer to the critical point where the mass vanishes. A more
complete analysis of this system which examines the costs and benefits
of keeping more states versus computing longer-range connected
contributions to the renormalized Hamiltonian would be very informative.

\section{Deriving the Basic Rules}
\label{sec:derive}

The definition of the renormalized Hamiltonian in Eq.~\ref{hamren}
is the cornerstone of the CORE approach.  We were led to this
definition by first observing that the state
 \begin{equation}
   \vert\Psi(t)\rangle ={\displaystyle e^{-tH/2} \over
                  \sqrt{\langle\Psi_0\vert\,
                   e^{-tH}\,\vert\Psi_0\rangle}}\vert\Psi_0\rangle
   \label{eqi}
 \end{equation}
contracts, as $t\rightarrow\infty$,
 onto the lowest energy eigenstate of $H$ for
which the starting trial state $\vert\Psi_0\rangle$
 has a nonvanishing overlap,
typically the ground state.   Note that the ground-state expectation value
of any operator $O$ can be obtained by taking the limit
 \begin{equation}
 \langle O \rangle =  \lim_{t \rightarrow \infty} \langle\Psi(t)\vert O
  \vert\Psi(t)\rangle.
   \label{eqlimit}
 \end{equation}
The use of Eq.~\ref{hamren} to define the renormalized Hamiltonian
is strongly suggested by the following theorem.

\begin{quote}
{\bf Theorem I:} Given a Hilbert space ${\cal H}$ and
truncation algorithm with associated projection operator $P$,
diagonalizing the renormalized Hamiltonian
 \begin{equation} H^{\rm ren}(t) =
[\! [\,T(t)^2\,]\! ]^{-\frac{1}{2}}\,[\! [\,T(t)\,H\,T(t)\,]\! ]
         \,[\! [\,T(t)^2\,]\! ]^{-\frac{1}{2}},
 \label{hamrent}
 \end{equation}
where $T(t)=e^{-tH}$ and $[\! [O]\! ]=P^\dagger OP$,
is equivalent to finding the vector $\vert\psi\rangle$ in the subspace
${\cal P}=P\,{\cal H}$ which minimizes the ratio
\begin{equation}
 {\cal E}_\psi(t) = \langle \psi\vert
T(t)\,H\,T(t)\vert\psi\rangle /\langle \psi\vert T(t)^2\vert\psi\rangle.
\end{equation}
\end{quote}

\begin{quote}
{\sl Proof:}  Let $\vert\phi_n\rangle$
label a basis for ${\cal P} = P\,{\cal H}$ and
expand $\vert\psi\rangle = \sum_n\, a_n \vert\phi_n\rangle$. Then,
 \begin{equation}
   {\cal E}_\psi(t) = {\sum_{n,m}\, a_m^{\ast} a_n
      \,\langle \phi_m\vert T(t)\,H\,T(t) \vert\phi_n\rangle \over
      \sum_{p,q}\,a_p^{\ast} a_q \langle \phi_p\vert
       T(t)^2 \vert\phi_q\rangle }.
   \label{GENvar}
 \end{equation}
To minimize this expression over all the states in ${\cal P}$,
differentiate Eq.~\ref{GENvar} with respect to $a_r^{\ast}$ and equate
to zero; this yields
\begin{equation}
 \langle \phi_r\vert T(t)\, H \, T(t)\vert \psi\rangle
 = {\cal E}_\psi(t) \langle \phi_r\vert T(t)^2\vert\psi\rangle.
\end{equation}
Since this applies for each $r$, this can be rewritten
\begin{equation}
    [\! [\,T(t)\,H\,T(t)\,]\! ]\, \vert\psi\rangle
      = {\cal E}_\psi(t) \,[\! [\,T(t)^2\,]\! ]\, \vert\psi\rangle.
   \label{GENdiff}
\end{equation}
Hence, finding the state $\vert\psi\rangle$ which minimizes
${\cal E}_\psi(t)$ is equivalent to solving a generalized eigenvalue
problem.  Given that $[\! [\,T(t)^2\,]\! ]$ is a positive matrix,
the relative eigenvalue problem can be converted
into an equivalent standard eigenvalue problem.  In other words,
finding the state which minimizes the ground-state energy of the
original Hamiltonian is equivalent to diagonalizing the
operator $H^{\rm ren}(t)$ defined in Eq.~\ref{hamrent}.
\end{quote}

In order to demonstrate that the renormalized Hamiltonian given
in Eq.~\ref{hamren} defines a valid renormalization group
transformation $H^{\rm ren}=\tau(H)$, we must at least show that
the eigenvalues of $H^{\rm ren}$ and the low-lying eigenvalues
of the original Hamiltonian $H$ are the same.

\begin{quote}
{\bf Theorem II:} Let $H$ be a Hamiltonian defined in an
$N$-dimensional Hilbert space ${\cal H}$ with eigenstates
$\{\vert\epsilon_\beta\rangle\}$ and corresponding eigenvalues
$\epsilon_\beta$, for $\beta=0\dots N-1$ and $\epsilon_\beta\leq
\epsilon_{\beta+1}$.  Let ${\cal P}$ be an $M$-dimensional subspace
${\cal P} \subset {\cal H}$ spanned by the states $\{\vert\phi_\alpha
\rangle\}$ for $\alpha=0\dots M-1$, where $M<N$.  The projector
into this subspace is $P=\sum_{\alpha=0}^{M-1} \vert\phi_\alpha\rangle
\langle\phi_\alpha\vert$.  Furthermore, let $S$ denote the $M\times M$
unitary triangulation matrix whose construction has been previously
described, and $\{\vert{\cal T}_\beta\rangle\}$ denote the remnant
eigenstates of $H$ in ${\cal P}$ with corresponding eigenvalues
${\cal T}_\beta$.  Then the operator defined by
\begin{equation}
 H^{\rm ren} = \lim_{t\rightarrow\infty}\
 [ \! [\, T(t)^2 \,] \! ]^{-1/2}\ [ \! [ T(t)\ H
  \ T(t) \,] \! ]\ [ \! [\, T(t)^2 \,] \! ]^{-1/2},
 \label{hamrdef}
\end{equation}
where the contractor $T(t)=e^{-tH}$ and $[\! [ O ]\! ]=P^\dagger O P$,
simplifies to
\begin{equation}
 H^{\rm ren} = S^\dagger H_{\cal T} S ,
 \label{hamrdefsimp}
 \end{equation}
where $H_{\cal T}={\rm diag}({\cal T}_0,\dots,{\cal T}_{M-1})$.
\end{quote}

\begin{quote}
{\sl Proof:}  Define the $M$ states $\vert\xi_\alpha\rangle$ by
\begin{equation}
    \vert\xi_\alpha\rangle = \sum_{\rho=0}^{M-1} \vert\phi_\rho\rangle
    \ S^\dagger_{\rho\alpha}.
\end{equation}
{}From the construction of $S$ and the remnant eigenstates using
the singular value decomposition, the states $\vert\xi_\alpha\rangle$
satisfy $\langle{\cal T}_\rho\vert\xi_\alpha\rangle =0$ for
$\rho<\alpha$.  However, since the projections of missed or
non-remnant eigenstates into ${\cal P}$ can be expressed in terms
of the projections of lower-lying remnant states, this means that
$\langle\epsilon_\rho\vert\xi_\alpha\rangle =0$ for all
$\rho$ satisfying $\epsilon_\rho < {\cal T}_\alpha$.
The use of Eq.~\ref{calTdef} and the singular value decomposition
to define the remnant eigenstates also ensures that
$\langle{\cal T}_\rho\vert\xi_\alpha\rangle=0$ even for $\rho$
corresponding to remnant states which are degenerate with but
orthogonal to $\vert{\cal T}_\alpha\rangle$, and hence,
\begin{equation}
  e^{-tH}\vert\xi_\alpha\rangle = e^{-t{\cal T}_\alpha}
  \vert{\cal T}_\alpha\rangle \langle{\cal T}_\alpha\vert
  \xi_\alpha\rangle +  \sum_{\epsilon_\beta
  > {\cal T}_\alpha} e^{-t\epsilon_\beta}\vert\epsilon_\beta\rangle
  \langle\epsilon_\beta\vert\xi_\alpha\rangle,
\end{equation}
where $\langle{\cal T}_\alpha\vert\xi_\alpha\rangle \neq 0$.
Now define a new set of states
\begin{eqnarray}
  \vert\Theta_\alpha(t)\rangle &=& Z_{\alpha\beta}(t)\ \vert
  \xi_\beta\rangle,\\
  Z_{\alpha\beta}(t) &=& \delta_{\alpha\beta}
  \ e^{t{\cal T}_\alpha} \vert\langle\xi_\alpha\vert{\cal T}_\alpha
  \rangle\vert^{-1}.
\end{eqnarray}
It is not difficult to show that
\begin{eqnarray}
  \lim_{t\rightarrow\infty} \langle\Theta_\alpha(t)\vert
 T(t)^2\vert\Theta_\beta(t)\rangle &=& \delta_{\alpha\beta},\\
  \lim_{t\rightarrow\infty}\langle\Theta_\alpha(t)\vert
 T(t) H T(t)\vert\Theta_\beta(t)\rangle &=& {\cal T}_\alpha
  \delta_{\alpha\beta}.
\end{eqnarray}
In matrix notation, these equations are
\begin{eqnarray}
  \lim_{t\rightarrow\infty} Z(t) S [\![ T(t)^2 ]\!] S^\dagger
  Z(t) &=& I, \label{srtinv}\\
  \lim_{t\rightarrow\infty} Z(t) S [\![ T(t)HT(t) ]\!] S^\dagger
  Z(t) &=& H_{\cal T} \label{tinflim},
\end{eqnarray}
where $I$ is the $M\times M$ identity matrix.  It then follows
from Eq.~\ref{srtinv} that
\begin{equation}
  \lim_{t\rightarrow\infty} \biggl( S [\![ T(t)^2 ]\!] S^\dagger
  \biggr)^{-\frac{1}{2}} = \lim_{t\rightarrow\infty} Z(t),
\end{equation}
and thus, Eq.~\ref{tinflim} becomes
\begin{equation}
  \lim_{t\rightarrow\infty} \biggl( S [\![ T(t)^2 ]\!] S^\dagger
  \biggr)^{-\frac{1}{2}} \biggl( S [\![ T(t)H T(t) ]\!] S^\dagger
  \biggr) \biggl( S [\![ T(t)^2 ]\!] S^\dagger
  \biggr)^{-\frac{1}{2}} = H_{\cal T}.
\end{equation}
Using the matrix relation $BA^{-1/2}B^{-1}=(BAB^{-1})^{-1/2}$
and the unitarity of $S$, it then follows that
\begin{equation}
  \lim_{t\rightarrow\infty}  S [\![ T(t)^2 ]\!]^{-\frac{1}{2}}
    [\![ T(t)H T(t) ]\!]  [\![ T(t)^2 ]\!]^{-\frac{1}{2}}
     S^\dagger = H_{\cal T},
\end{equation}
and finally,
\begin{equation}
 H^{\rm ren} = S^\dagger H_{\cal T} S.
\end{equation}
\end{quote}

This theorem demonstrates that the eigenvalues of the
renormalized Hamiltonian are the $M$ eigenvalues ${\cal T}_\alpha$
associated with the remnant eigenstates of the original
Hamiltonian.  If the truncation procedure is such that no eigenvalues
are missed, then the eigenvalues of $H^{\rm ren}$ are the $M$ lowest
eigenvalues of $H$.  By showing that the mapping $\tau(H)=H^{\rm ren}$
replaces the original theory with a theory containing the same
low-energy physics but defined in terms of fewer degrees of freedom,
Theorem II provides the justification for identifying $\tau$
as a {\em renormalization group} transformation.

An important aspect of the CORE approach is the use of the finite
cluster method to approximate the renormalized Hamiltonian on the
infinite lattice.  In Ref.~\cite{Sykes}, lattice constant theory was
used to show that the finite cluster method can be applied in the
calculation of any quantity so long as that quantity is extensive.
We now demonstrate the extensivity of the renormalized Hamiltonian.
Recall that a quantity is extensive if, when evaluated on a
disconnected graph, it is the sum of that quantity evaluated
separately on the connected components of the graph.

\begin{quote}
{\bf Theorem III:}
 The renormalized Hamiltonian is extensive.
 \end{quote}

\begin{quote}
 {\sl Proof:}
Consider a disconnected sublattice $G=G_1 \cup G_2$ comprised of two
connected components $G_1$ and $G_2$.  Since $G$ is disconnected,
$H(G)=H(G_1)+H(G_2)$ and
$[H(G_1),H(G_2)]=0$; hence, $T_G(t)=T_{G_1}(t)T_{G_2}(t)=
T_{G_2}(t)T_{G_1}(t)$.  Since the truncation is done on a
block-by-block basis, then
$[\![T_{G}(t)]\!] = [\![T_{G_1}(t)]\!]\ [\![T_{G_2}(t)]\!]$ and
\begin{eqnarray}
  H^{\rm r}(G) &=& [\![T_{G_1}(t)^2]\!]^{-1/2}
 [\![T_{G_2}(t)^2]\!]^{-1/2}\biggl( [\![T_{G_1}(t)H(G_1)T_{G_1}(t)]\!]
 [\![T_{G_2}(t)^2]\!] \nonumber\\
 & + & [\![T_{G_2}(t)H(G_2)T_{G_2}(t)]\!] [\![T_{G_1}(t)^2]\!]
 \biggr) [\![T_{G_1}(t)^2]\!]^{-1/2}
 [\![T_{G_2}(t)^2]\!]^{-1/2}, \\
&=& [\![T_{G_1}(t)^2]\!]^{-1/2} [\![T_{G_1}(t)H(G_1)T_{G_1}(t)]\!]
 [\![T_{G_1}(t)^2]\!]^{-1/2} \nonumber\\
& +& [\![T_{G_2}(t)^2]\!]^{-1/2} [\![T_{G_2}(t)H(G_2)T_{G_2}(t)]\!]
 [\![T_{G_2}(t)^2]\!]^{-1/2}, \\
&=& H^{\rm r}(G_1) + H^{\rm r}(G_2).
\end{eqnarray}
Hence, $H^{\rm r}$ is extensive.
\end{quote}

Clearly, the eigenvalues of the renormalized Hamiltonian on a
given cluster containing $R$ blocks are the same as the lowest
$M^R$ eigenvalues (modulo missing ones) of the full Hamiltonian
restricted to the cluster, assuming $M$ states are retained
in each block.  In truncating the cluster expansion of the
renormalized Hamiltonian on the infinite lattice, the
correspondence between the low-lying eigenvalues of the
infinite lattice $H^{\rm ren}$ and $H$ can then be only
approximate.  However, our previous examples suggest that
truncating the cluster expansion of the renormalized Hamiltonian
after only a very few terms can lead to remarkably accurate
results.

\section{Approximation Issues}
\label{sec:converge}

In this section, we discuss two issues related to approximations and
the CORE procedure.  First, within the context of free scalar field
theory, we link the methods presented in this paper to our earlier
Physical Review letter where we used approximate contractors to carry
out the computations.  We demonstrate how such approximations converge
and show why previous approaches always found a best finite value of
$t$ for determining the ground-state energy.  Second, we reconsider the
question of single-state truncations in the Heisenberg antiferromagnet.
We do this to show how simple single-state truncations can encounter
problems with surface effects and how the multistate renormalization
group algorithm avoids these problems.

\subsection{Connection To Earlier Methods: Approximate Contractors}
\label{Tnapprox}

Our earlier version of the CORE procedure \cite{COREPrl}
used an approximate contractor $T_n(t)$ obtained by
decomposing $H= H_1 + H_2$ into two or more parts and writing
 \begin{equation}
   T_n(t) = \left[\,e^{-t H_1/2n}\,e^{-t H_2/n}\,e^{-t H_1/2n}
   \,\right]^n,
   \label{iii}
 \end{equation}
where $H_1$ and $H_2$ are chosen such that $e^{-t H_1}$ and
$e^{-t H_2}$ could either be computed exactly or numerically
to any desired degree of accuracy.  The validity of this approximation
follows from the fact that, for operators $A$ and $B$, one can show that
 \begin{eqnarray}
      e^{\delta (A+B)} &=& e^{\delta A/2}\,e^{\delta B/2}\,e^{C(\delta)}
         \,e^{\delta B/2}\,e^{\delta A/2}, \\
      C(\delta) &=& \sum_{j=1}^{\infty} \delta^{2j+1} O_{2j+1}.
   \label{trotter}
 \end{eqnarray}
In particular, for $\delta = t/n$ we see that as $n \rightarrow \infty$,
the sequence $T_n(t)$ converges to $e^{-tH}$ as $(t/n)^3$.  When
approximating the contractor in this way, we will see that $t$ must
be viewed as a variational parameter to be optimized.  We will also
see that this earlier procedure is less accurate and more time
consuming than the method presently proposed.

To see this, consider once again the Hamiltonian given in
Eq.~\ref{bosefreefield} which describes a free scalar field theory
in one spatial dimension.  This Hamiltonian can be expressed as a
sum of single-site operators and nearest-neighbor interactions:
$H=H_0+V$, where
 \begin{eqnarray}
      H_0 &=&  \sum_j  \frac{1}{2} \left[ \Pi(j)^2 +
       \gamma_0^2\ \phi(j)^2 \right], \\
      V &=& - \sum_j \phi(j)\phi(j+1),
   \label{BFFHdecomp}
 \end{eqnarray}
and $\gamma_0 = \sqrt{\mu^2+2}$.  The ground state of $H_0$
is then a product of uncorrelated Gaussians:
 \begin{equation}
   \langle\phi\vert\Psi_0\rangle = \prod_j \langle\phi
   \vert\gamma_0(j)\rangle = \prod_j e^{-\gamma_0 \phi(j)^2/2}.
   \label{BFFtrialstate}
 \end{equation}
Our aim is now two-fold: to demonstrate how to recover the
ground state of the full theory by applying $e^{-tH}$ to
$\vert\Psi_0\rangle$ and taking the limit
$t\rightarrow\infty$; and to determine how well
$T_n(t) = T_1(t/n)^n = [e^{-tH_0/2n} e^{-tV/n} e^{-tH_0/2n}]^n$
approximates $e^{-tH}$ for finite values of $n$ and $t$.

First, evaluate $e^{-tH}\vert\Psi_0\rangle$. Introducing the Fourier
transforms $\tilde\phi(k)$ and $\tilde\Pi(k)$, we obtain
 \begin{eqnarray}
  H &=&  \sum_{k}\, H(k) = \sum_k\,
      \frac{1}{2}\left[  \tilde\Pi(-k)\tilde\Pi(k)
       + \omega^2(k)\ \tilde\phi(-k)\tilde\phi(k) \right], \\
  H_0 &=&  \sum_k\, H_0(k) = \sum_{k}\,
      \frac{1}{2} \left[ \tilde\Pi(-k)\tilde\Pi(k)
       + \gamma_0^2\ \tilde\phi(-k)\tilde\phi(k) \right], \\
  V &=&  \sum_k\, V(k) = -\sum_{k}\,
       \cos(k)\, \tilde\phi(-k)
       \tilde\phi(k), \\
  \langle\tilde\phi\vert\Psi_0\rangle &=& \prod_k \exp\left[- \gamma_0
       \,\tilde\phi(-k)\tilde\phi(k) /2 \right],
   \label{BFFfforms}
 \end{eqnarray}
where $\omega(k) = \sqrt{\mu^2 + 4\sin^2(k/2)}$.
Since $\tilde\phi(-k) = \tilde\phi(k)^{\dag}$ and
$\tilde\Pi(-k)=\tilde\Pi(k)^{\dag}$, we can decompose the
fields in terms of their real and imaginary parts and
restrict all sums to $k > 0$, handling the case $k=0$ separately.
Since the $H(k)$ mutually commute, then $e^{-tH}\vert\Psi_0\rangle$
can be written as a product over states labelled by the
momentum $k$ so that we can limit our attention to a single
$k$-mode without loss of generality.
Let $\vert \gamma(k)\rangle$ denote a simple harmonic
oscillator ground state of frequency $\gamma(k)$.
Now apply Theorem A2 proven in Appendix~\ref{app:sho}:
 \begin{equation}
   e^{-t\,H(k)} \vert\gamma(k)\rangle
  = A(k,t)\vert\gamma(k,t)\rangle,
   \label{BFFgammat}
 \end{equation}
where $A(k,t)$ is a normalization factor and
$\vert\gamma(k,t)\rangle$ is a simple harmonic oscillator
state of frequency
 \begin{equation}
 \gamma(k,t) = \omega(k)\, \left(\frac{
   \gamma_0+\omega(k) + e^{-2\,t\,\omega(k)} (\gamma_0-\omega(k))}{
   \gamma_0+\omega(k) - e^{-2\,t\,\omega(k)} (\gamma_0-\omega(k))}
   \right).
   \label{BFFgammaoft}
 \end{equation}
Thus, as $t\rightarrow\infty$, the frequencies $\gamma(k,t)\rightarrow
\omega(k)$ which means that
the state $e^{-tH} \vert\Psi_0\rangle$ converges (up to
the normalization factor $\prod_k A(k,t)$) to the true ground state of
the lattice free field theory.  Since the normalization factor cancels
out in ratios such as
 \begin{equation}
   {\langle \Psi_0\vert e^{-t H/2} H e^{-t H/2} \vert\Psi_0\rangle \over
      \langle \Psi_0\vert e^{-t H} \vert\Psi_0\rangle }
         = \frac{1}{2} \sum_{k\geq 0}\,\left( \gamma(k,t) +
         \frac{\omega^2(k)}{\gamma(k,t)} \right),
   \label{BFFfreeenergy}
 \end{equation}
we will ignore it from here on.

The determination of $T_n(t)\vert\Psi_0\rangle$ proceeds similarly
to that of $e^{-tH}\vert\Psi_0\rangle$.  Since we have
$[H_0(k),H_0(k^\prime)]=0$, $[V(k),V(k^\prime)]=0$
and $[H_0(k),V(k^\prime)]=0$ for $k\neq k^\prime$, then
$T_n(t) \vert\Psi_0\rangle$ can be written as a
product over states labelled by the variable $k$ and we can study the
general problem one $k$-mode at a time.   If $\vert\gamma_p(k)\rangle$
is a simple harmonic oscillator ground state of frequency $\gamma_p(k)$
associated with a mode $k$, then
\begin{eqnarray}
 e^{-tH_0(k)/2n}\,\vert\gamma_p(k)\rangle &\propto&
   \vert\gamma^\prime(k,t/2n)\rangle,\\
 e^{-tV(k)/n}\,\vert\gamma_p(k)\rangle &\propto&
  \vert\gamma^{\prime\prime}(k,t/n) \rangle,
\end{eqnarray}
where the frequencies of the new oscillator
ground states are related to $\gamma_p(k)$ by
\begin{eqnarray}
   \gamma^\prime(k,t) &=& \gamma_0\, \left(\frac{
       \gamma_p(k)+\gamma_0 + e^{-2\,t\,\gamma_0}
        (\gamma_p(k)-\gamma_0)}{
       \gamma_p(k)+\gamma_0 - e^{-2\,t\,\gamma_0}
        (\gamma_p(k)-\gamma_0)}
       \right),\\
   \gamma^{\prime\prime}(k,t) &=& \gamma_p(k)-2t\cos(k).
\end{eqnarray}
Using the above relations,
$[e^{-tH_0(k)/2n}e^{-tV(k)/n}e^{-tH_0(k)/2n}]^n\,
\vert\gamma_0(k)\rangle$ can then be easily evaluated. We find that
the Gaussian state $\vert\gamma_0(k)\rangle$ evolves to a new Gaussian
state $\vert\gamma_n(k,t)\rangle$ of frequency $\gamma_n(k,t)$.

\begin{figure}
\begin{center}
\leavevmode
\epsfxsize=5in\epsfbox[80 160 530 760]{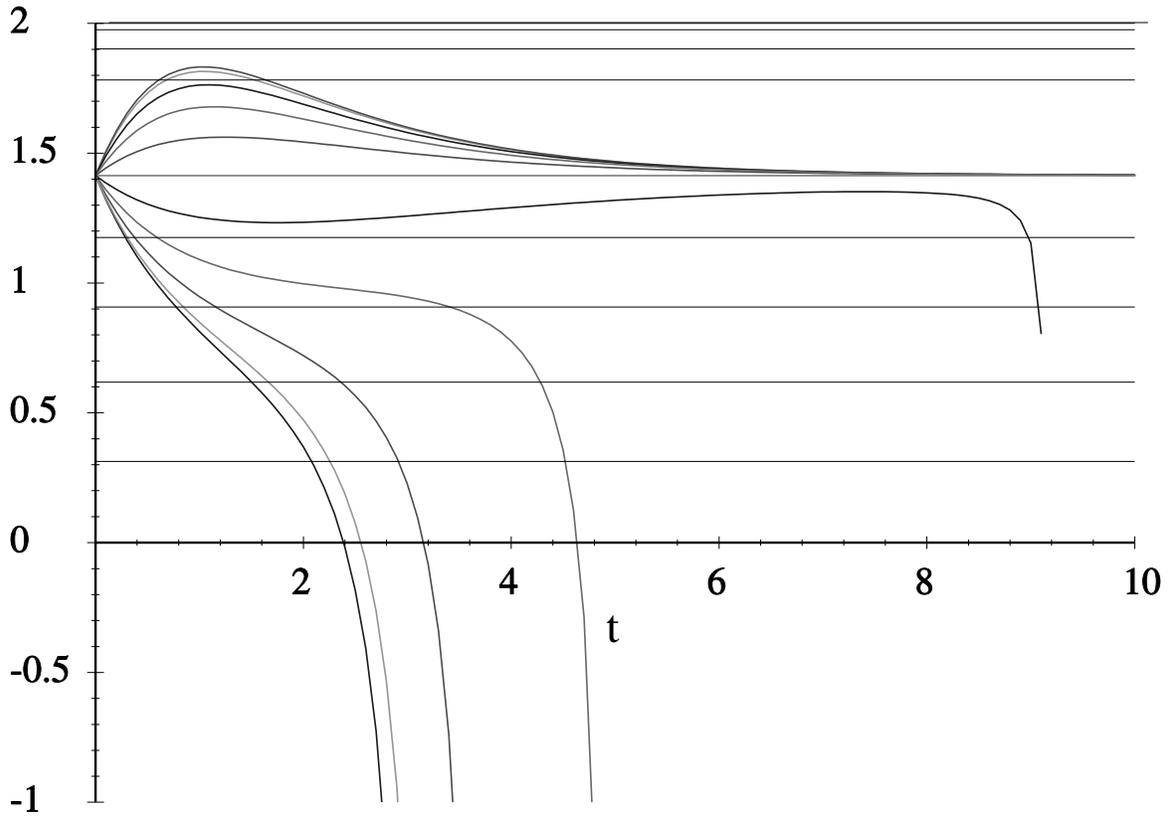}
\end{center}
\caption[freefield]{The scalar field theory frequencies
$\gamma_n(k,t)$ obtained using an approximate contractor $T_n(t)$
for $n=1$ and various momenta $k$.  The starting state is a
product of uncorrelated Gaussians.  The curves correspond to
different values of $k$; the corresponding frequencies in
the exact wavefunction are indicated by the horizontal lines.}
\label{FREEFIELDONE}
\end{figure}

\begin{figure}
\begin{center}
\leavevmode
\epsfxsize=5in\epsfbox[80 160 530 760]{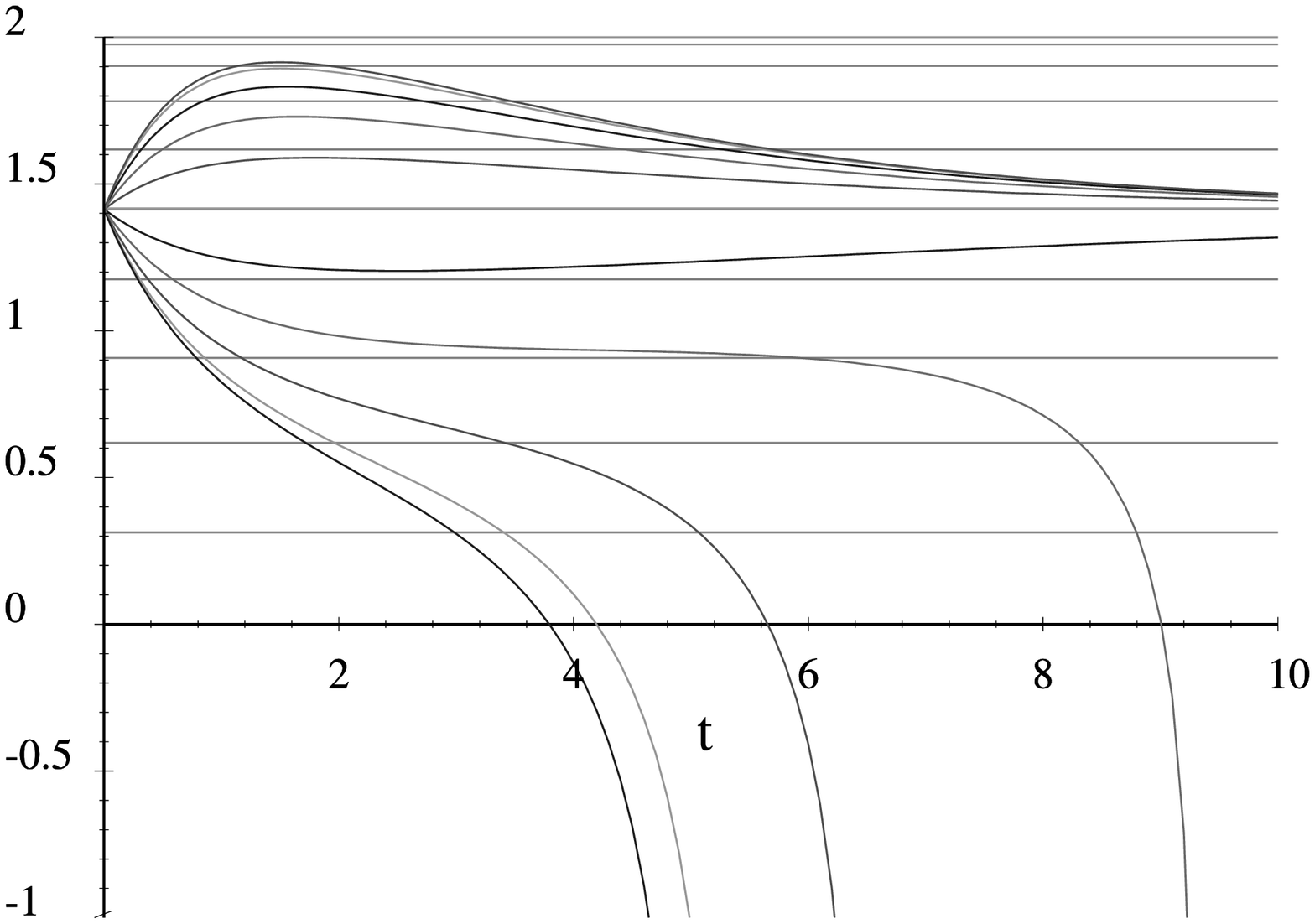}
\end{center}
\caption[freefieldtwo]{The frequencies $\gamma_n(k,t)$ as
in Fig.~\ref{FREEFIELDONE}, except that $n=2$.}
\label{FREEFIELDTWO}
\end{figure}

\begin{figure}
\begin{center}
\leavevmode
\epsfxsize=5in\epsfbox[80 160 530 760]{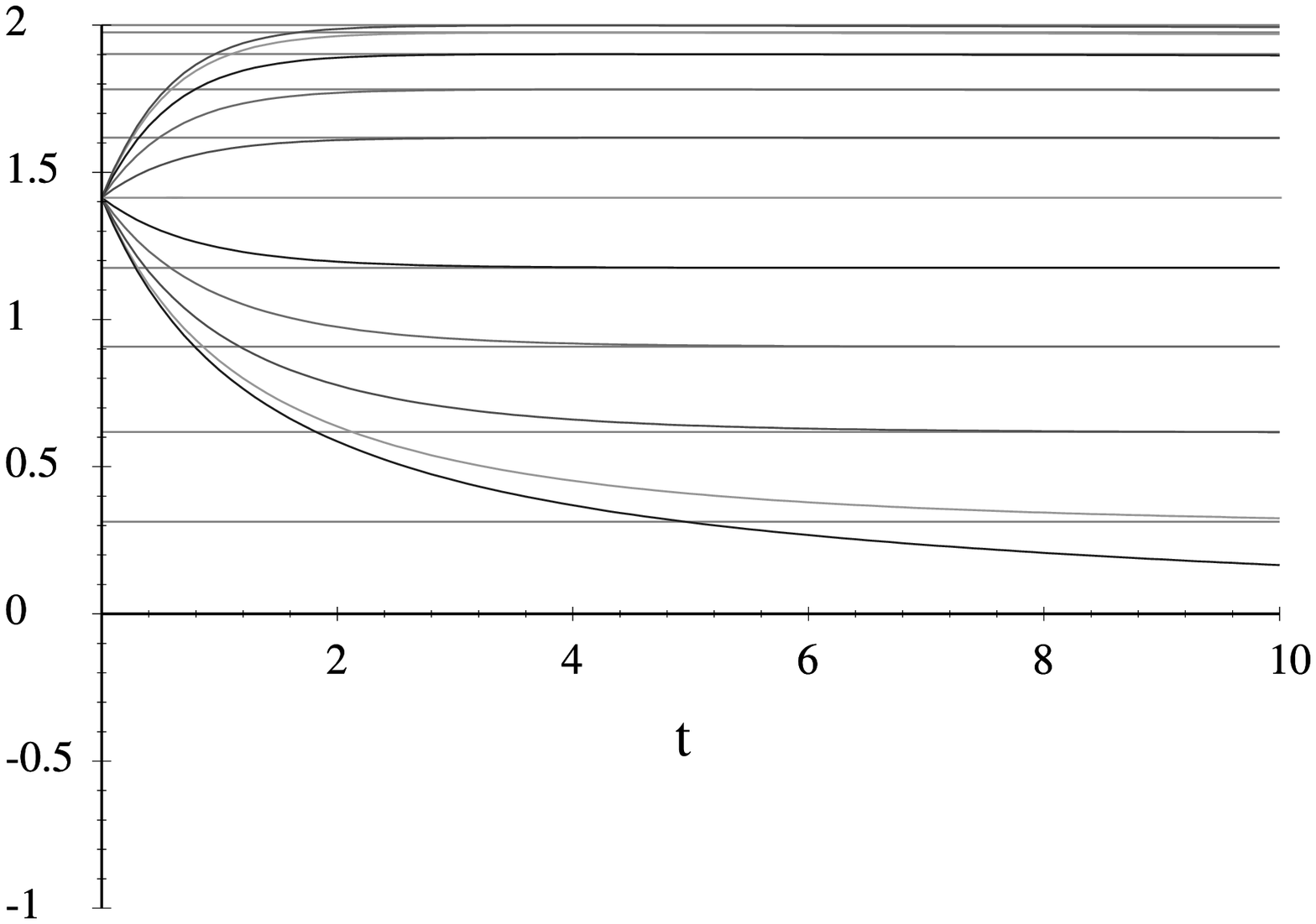}
\end{center}
\caption[freefieldforty]{The frequencies $\gamma_n(k,t)$ as
in Fig.~\ref{FREEFIELDONE}, except that $n=40$.}
\label{FREEFIELDFORTY}
\end{figure}

\begin{figure}
\begin{center}
\leavevmode
\epsfxsize=5in\epsfbox[80 160 530 760]{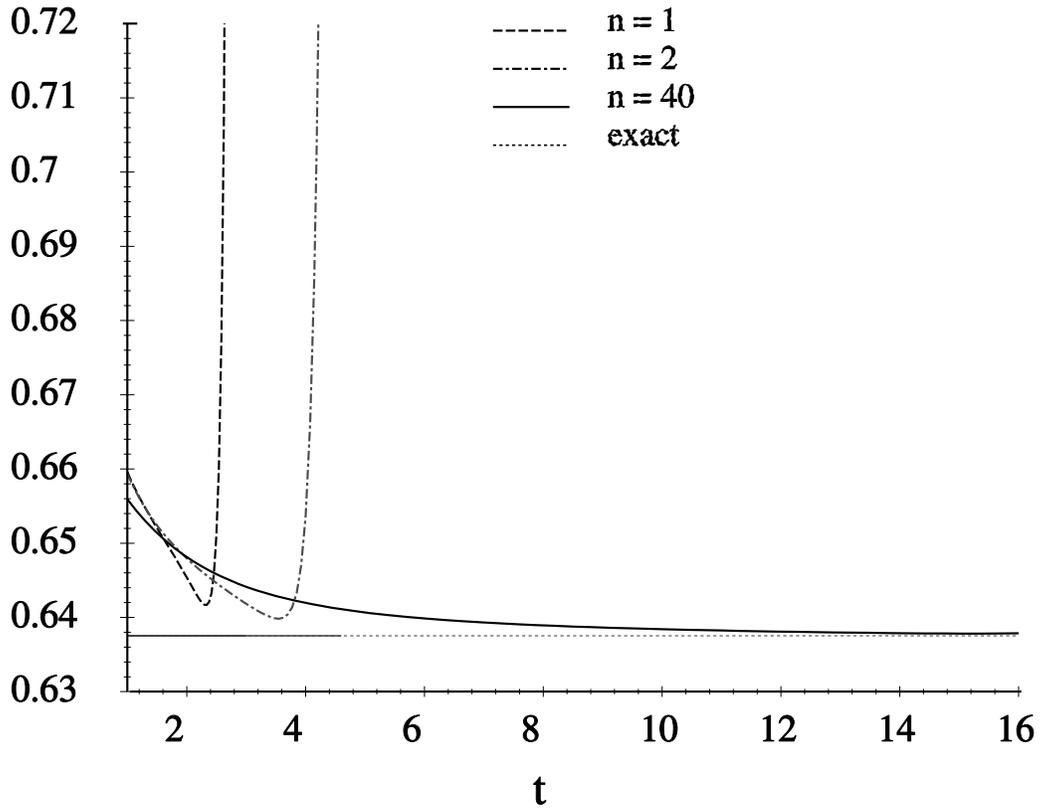}
\end{center}
\caption[eoftone]{Expectation value of the free scalar field theory
Hamiltonian in the state $T_n(t)\,\vert\Psi_0\rangle$
for $n=1,2$, and $40$, showing how the
best $t$ changes as a function of $n$. }
\label{EOFTFG}
\end{figure}

Plots of $\gamma_n(k,t)$ for various values of $k$ and $n$ and a range
of $t$ values are shown in Figs.~\ref{FREEFIELDONE}, \ref{FREEFIELDTWO}
and \ref{FREEFIELDFORTY}.  Plots of the expectation value of $H$ in the
state $T_n(t) \vert\Psi_0\rangle$ for the same values of $n$ and range
of $t$ are shown in Fig.~\ref{EOFTFG}. There are two things to notice
about these figures. First, for larger values of $k$,
the frequencies converge quickly to the values they would have in the
exact wavefunction, indicated by the horizontal lines; however, for
smaller values of $k$, the exact frequencies are not well reproduced,
even for very large values of $n$.  This means that computing the
action of $T_n(t)$ on $\vert\Psi_0\rangle$ can do well at approximating
the ground-state energy density and still fail to reproduce the mass
gap. Second, we observe that for finite values of $n$, there is a
finite $t$ which yields a best estimate of the ground-state energy
density.  $T_n(t)$ does a very good job of approximating $e^{-tH}$ for
smaller values of $t$, so at first, $T_n(t) \vert\Psi_0\rangle$ tends
towards the ground state of $H$;  however, $T_n(t)\vert\Psi_0\rangle$
eventually begins to move away from the ground-state wavefunction and
so the expectation value of the energy density starts to get worse.
This shows that without additional improvements, working with $T_n(t)$
for finite $n$ and a {\em best} $t$ cannot be expected to always
accurately reconstruct the infrared properties of the theory.  The
renormalization group method works better than simply evaluating the
action of $T_n(t)$ on a single state because it eliminates only the
higher states which $T_n(t)$ reproduces well and carries the more
difficult long-wavelength modes over to the next step of the
calculation.  The agreement between the results of our earlier CORE
treatment of the Ising model which used $n \le 16$ and a best value of
$t$ and our current $n=\infty$ and $t=\infty$ calculation supports this
picture.

\subsection{Antiferromagnet: Simple Cluster Formulae}

We now return to the Heisenberg antiferromagnet and compute the
vacuum energy density using two different
single-state truncation procedures.
There are two reasons for doing this: first, to show that computing
the ground-state energy density for an infinite-volume theory from
a series of finite-volume calculations is generally applicable;
second, we often learn more from examples which do not work as
expected than from ones which work well.  In this case, we will
learn that partitioning the lattice into either two- or three-site
blocks can produce sequences of truncated cluster expansions which
converge at very different rates.  We explain why this happens
and show how the two-state truncation algorithm used earlier avoids
these convergence problems.

First, we apply a single-state RG algorithm in which the lattice is
partitioned into two-site blocks and we retain only the lowest-lying
eigenstate in each block.  Denote by $E_r$ the ground-state energy of
the theory defined by restricting the full Hamiltonian to an $r$-site
sublattice. The two-, four-, six-, and eight-site ground-state energies
are $E_2 = -0.75$, $E_4 = -1.616025$, $E_6 = -2.493577$, and $E_8 =
-3.374932$, and they yield the following connected contributions in the
cluster expansion of the renormalized Hamiltonian:
 \begin{eqnarray}
      \epsilon_2 &=& E_2 = -0.75, \\
      \epsilon_4 &=& E_4 - 2 \epsilon_2 = -0.116025, \\
      \epsilon_6 &=& E_6 - 2 \epsilon_4 - 3 \epsilon_2 = -0.011527, \\
      \epsilon_8 &=& E_8 - 2 \epsilon_6 - 3 \epsilon_4 - 4 \epsilon_2 =
          -0.003803.
   \label{FRCffclusttwosite}
 \end{eqnarray}
Thus, we obtain a sequence of approximations to the infinite-volume
ground-state energy density from the following truncated
cluster expansions:
 \begin{eqnarray}
      {\cal E}_2 &=& \epsilon_2/2 = -0.375, \\
      {\cal E}_{24} &=& (\epsilon_2+\epsilon_4)/ 2 = -0.4330125, \\
      {\cal E}_{246} &=& (\epsilon_2+\epsilon_4+\epsilon_6)/ 2
            = -0.438776, \\
      {\cal E}_{2468} &=& (\epsilon_2+\epsilon_4+\epsilon_6 + \epsilon_8)
          /2 = -0.4406775,
  \end{eqnarray}
which are to be compared to the exact energy density ${\cal E}_{\rm
exact} =  -0.443147$. Note that we divide by two in the above formulas
so that our results refer to the energy per site of the original lattice
instead of the energy per two-site block. For this simple truncation
algorithm, the finite-range cluster expansion converges rapidly and
agreement with the exact answer to better than one percent is obtained
with ease.  Given our earlier discussion of the free-field theory, it
is interesting to compare the approximations built from connected terms
to what we would obtain from simply dividing the ground-state energy
for each $n$-site block by $n$.  The comparison of these results is
presented in Table~\ref{haftable}.

The better than one-percent agreement of the finite-range cluster
expansion with the exact ground-state energy density brings into
question the benefits of using the renormalization group algorithm.
However, the need for the renormalization group becomes apparent
after examining the sequence of approximations obtained using
three-site blocks.  In this case, numerical diagonalization of the
appropriate sublattice Hamiltonians yields $E_3 = -1.0$,
$E_6 = -2.493577$, and $E_9 = -3.736322$, yielding connected
contributions
 \begin{eqnarray}
      \epsilon_3 &=& E_3 = -1.0, \\
      \epsilon_6 &=& E_6 - 2 \epsilon_3 = -0.493577, \\
      \epsilon_9 &=& E_9 - 2 \epsilon_6 - 3\epsilon_3 = 0.250832.
   \label{FRCffclustthreesite}
 \end{eqnarray}
If we now use these results to construct the corresponding
approximations to the energy density per site, we obtain
the sequence
 \begin{eqnarray}
      {\cal E}_3 &=& \epsilon_3/3 = - 1/3, \\
      {\cal E}_{36} &=& (\epsilon_3 + \epsilon_6)/ 3 = -0.497859, \\
      {\cal E}_{369} &=& (\epsilon_3 + \epsilon_6 + \epsilon_9)
               /3 = -0.4142483,
  \end{eqnarray}
which oscillates about the correct answer and converges much
more slowly than that for the two-site decomposition of $H$.
The cause of this oscillation and slow convergence arises from the
fact that the physical excitations of this model have integer spin;
the three-site decomposition has difficulty reproducing the
low-lying physics since the ground state of the three-site block
is a spin-1/2 multiplet, that of the six-site block is spin-0,
and the ground-state of the nine-site block is once again spin-1/2.
The two-site decomposition of the Hamiltonian does not suffer from
this effect.  This lack of rapid convergence is very instructive;
since there is no way to know in advance what the correct
spectrum of excitations is, this shows that we need a method
for summing, at least partially, an infinite number of terms in
the finite-range cluster expansion.  As we saw in our earlier
discussion of the antiferromagnet, this is what the full
renormalization group calculation allows us to do.

\section{Looking Ahead}
\label{sec:looking}

This paper sets forth the basic rules for CORE computations, derives
the rules from first principles, and discusses issues related to the
convergence of the procedure.  Future papers will focus on the
application of these methods to more interesting physical systems and
on clarifying the connection of the CORE approach to perturbative
methods in instances where both are applicable.
Some systems which should receive early
attention are lattice gauge theories with and without fermions, t-J
models\cite{tJmodels}, and extended Hubbard models\cite{Hubbardmodels}.
It is important to study the application of CORE technology to lattice
gauge theories in order to see if, as we believe, it provides a
powerful alternative to Monte Carlo calculations for studying QCD and
chiral symmetry breaking. Extended Hubbard and t-J models are of
interest because they are conjectured to have some relevance to
high-$T_c$ superconductivity and have proven difficult to study in more
than one spatial dimension by conventional methods.  In this section,
we discuss the application of CORE methods to these problems and
indicate how one could establish the connection between the
CORE approach and a perturbative renormalization group treatment of
$\phi^4$ theory.

\subsection{Lattice Gauge Theory Without Fermions}

There are many ways to apply the techniques introduced in this paper to
lattice gauge theories. One interesting approach is to divide the
lattice into finite-size blocks, truncate the Hilbert space associated
with each block to a set of gauge-invariant states, and then use the
renormalization group formalism to map the gauge theory into a system
which, like a spin system, has only a finite number of states
associated with each lattice site.  This approach yields an
``equivalent'' Hamiltonian theory in which all of the unphysical
degrees of freedom have been eliminated.  We can then treat the new
Hamiltonian in the same way as in the Heisenberg and Ising models.

For example, we could associate with each {\em plaquette} of the
original lattice a single {\em site} in the new lattice.
We could then find the low-lying gauge-invariant eigenstates of
the one-plaquette Hamiltonian, either exactly or numerically, and
truncate by selecting a finite number of these eigenstates.  Using
this truncation procedure, we construct a renormalization group
transformation which maps the gauge theory into a generalized ``spin''
system.  The interactions between nearby ``spins'' are found by
evaluating the renormalized Hamiltonian on clusters containing
several connected plaquettes.  This new spin system would be
guaranteed to have the same low-lying gauge-invariant physics as
the original theory and could be treated in the same way as the
Heisenberg and Ising models.  This approach allows us to define
and carry out a gauge-invariant renormalization group calculation
for any lattice gauge theory.

This ability to define a gauge-invariant, Hamiltonian-based, real-space
renormalization group calculation is unique to the CORE approach.
Earlier real-space renormalization group procedures also kept a finite
number of states per block, but they defined the renormalized
Hamiltonian by $[[H]]$, the truncation of the original Hamiltonian
to the subspace spanned by the retained states (this corresponds to the
$t=0$ limit of the CORE approach).  In such calculations, keeping only
gauge-invariant block states leads to a truncated Hamiltonian in which
the block-block interactions vanish.  In order to retain inter-block
couplings, flux must move across the links joining the blocks; this
cannot happen without keeping some gauge-noninvariant single-block
states.  However, if one keeps such states in the truncation procedure,
the entire process becomes much more cumbersome.

The question of how many single-block gauge-invariant states and
how many terms in the cluster expansion of the renormalized Hamiltonian
should be retained naturally arises when carrying out a contractor
renormalization group calculation; each choice constructs a mapping
of the original gauge theory into a different generalized spin system.
We hope to answer this question in the future by carrying out
several computations in a simple lattice gauge theory, such as
2+1-dimensional compact $U(1)$, varying the number of retained
single-block states and clusters to see how quantities of interest,
such as mass gaps and the specific heat, depend on these factors.

\subsection{Lattice Gauge Theory With Fermions}

Interesting possibilities arise when we consider lattice gauge theories
with fermions.  One way of treating these theories is to study systems
with either SLAC\cite{slacderiv}, Wilson\cite{wilsonferm}, or
Quinn-Weinstein\cite{QuinnWein} fermions and truncate the system to the
subspace spanned by tensor products of gauge-invariant, single-site
states.  In the case of lattice QCD, this would include all
color-singlet single-site states, {\it i.e.,} mesons and baryons, which
can be formed by applying quark and antiquark creation operators to the
single-site vacuum state, subject to the constraints imposed by the
exclusion principle.  As the only terms which appear in the lattice QCD
Hamiltonian create (or destroy) closed loops of flux or move quarks
from site to site trailing their flux behind them, the color-singlet
mesons and baryons are all degenerate and the connected range-1 part of
the renormalized QCD Hamiltonian will vanish.  In order to compute the
connected range-2 terms, we solve the problem of two sites connected by
a single link and find the low-lying gauge-invariant eigenstates which
have an overlap with all of the tensor products of the two sets of
single-site meson and baryon states. This computation yields connected
range-2 contributions to the renormalized Hamiltonian which contain
meson and baryon kinetic terms as well as meson-meson and meson-baryon
interactions.  Connected range-3 terms come from computations involving
three sites arranged in a straight line or forming a right angle.
These range-3 terms contain corrections to the terms already described,
new terms which allow mesons and baryons to hop along diagonals of the
underlying lattice, and terms which describe three-site interactions.
Continuing in this way produces a renormalized Hamiltonian expressed
only in terms of the physical degrees of freedom; the underlying quarks
and gluons disappear from the problem.

We would now like to say something about how chiral symmetry
breaking will show up in QCD with three flavors of quarks
and either SLAC or Quinn-Weinstein fermion derivatives (the case of
Wilson fermions is somewhat different).
Consider a theory with three massless flavors of quarks and apply
a more restrictive truncation procedure which keeps only
single-site fluxless states containing equal numbers of quarks and
antiquarks, {\it i.e.,} mesons.  For three flavors of quarks there are
$924$ such states and, as was shown in Ref.~\cite{chiralsymmba},
they form an irreducible representation of the group $SU(12)$ where
the group generators are formed from bilinears in the single-site
quark fields $\vec{Q}$.  Note that for $N_f$ flavors, the fluxless
states form an irreducible representation of the group $SU(4N_f)$.
For a truncation algorithm based upon keeping gauge-invariant
single-site states, the renormalized Hamiltonian contains no range-1
connected terms.  The first nonvanishing contribution to the
renormalized Hamiltonian will be the range-2 connected terms and these
are computed by solving the two-site theory.

It was pointed out in Ref.~\cite{chiralsymmba} that if we keep only
the nearest-neighbor terms in the fermion derivative, then the
resulting Hamiltonian is invariant under a global $SU(12)$, and since
the two-site problem cannot have anything but nearest-neighbor terms,
this observation can be used to simplify the computation of the
connected range-2 terms in the renormalized Hamiltonian.  We already
noted that the fluxless single-site states form an irreducible
$924$-dimensional representation of $SU(12)$ and so tensor products
formed from these states can be decomposed into the irreducible
representations of $SU(12)$ which appear in the product of two $924$'s;
these are the only states in the full problem relevant to our CORE
computation.  Starting from the highest weight state in each of these
irreducible representations and applying the Lanczos method, we can
numerically find the relevant eigenvalues of the two-site Hamiltonian
to a high degree of accuracy. From general symmetry arguments, the most
general two-site Hamiltonian one can write for this system will be in
the form of a finite polynomial in the Casimir operator and higher
order invariants formed out of the generators of $SU(12)$.
Thus, the general structure of the connected range-2 Hamiltonian will
be given by
\begin{equation}
   h_2^{\rm conn}(j) = \alpha_1 \vec{Q}(j)\cdot\vec{Q}(j+1) +
      \alpha_2 (\vec{Q}(j)\cdot\vec{Q}(j+1))^2 + \ldots.
   \label{rangetwoqcd}
\end{equation}
It is a simple exercise to show that in the strong-coupling limit, the
leading term in this expansion is the one proportional to
$\vec{Q}(j)\cdot\vec{Q}(j+1)$; in other words, in strong-coupling, the
renormalized range-2 Hamiltonian is just a generalized Heisenberg
antiferromagnet.  As was argued in Ref.~\cite{chiralsymmba} and
Ref.~\cite{chiralsymmbb}, we expect this theory to spontaneously
break to $SU_V(6)\times SU_A(6)$, where the vector $SU_V(6)$ is realized
normally and the axial-vector $SU_A(6)$ is realized in the Goldstone
mode.  Thus, in the strong-coupling limit, the connected range-2 part of
the renormalized Hamiltonian unavoidably leads to a spontaneously
broken symmetry, but the group is too large and there are too many
Goldstone bosons.  Clearly, a detailed calculation is necessary
to determine if these conclusions persist in weak coupling where
other terms in Eq.~\ref{rangetwoqcd} can become significant.  However,
we can show that the problems of having too large a symmetry group
and too many Goldstone bosons disappears once we compute the
connected range-3 terms.

To see this, observe that, independent of the
coupling constant, the next-to-nearest-neighbor terms in both the SLAC
and Quinn-Weinstein types of derivative break the $SU(12)$ symmetry and,
after including these terms in the renormalized Hamiltonian, all that
remains of the $SU(6)\times SU(6)$ symmetry of the nearest-neighbor
theory is $SU(3)\times SU(3)$.  As in the discussion of the range-2
terms, we can invoke the strong-coupling limit to calculate the
structure of the leading range-3 terms and explicitly show that the
range-3 terms give the unwanted Goldstone bosons mass and that the
degenerate $SU(6)$ multiplets of mesons break up into $SU(3)$
multiplets.  This is in strict analogy to what was discussed in
Ref.~\cite{chiralsymmbb}.  Of course, as we noted for the case of the
range-2 terms, the generic structure of the connected range-3 terms in
the renormalized Hamiltonian is richer than that of the
leading terms in the strong coupling limit, and so asserting that this
pattern of symmetry breaking persists to the physically more
interesting weak-coupling regime requires more work than we have done
to this point.

Much interesting work remains to be done in this picture of
dynamical chiral symmetry breaking; nevertheless, the fact that
the CORE procedure provides a coupling-independent way of constructing
an effective theory of mesons which, in the strong-coupling limit,
coincides with earlier descriptions in which dynamical chiral
symmetry breaking appears naturally, is new and unique to this
approach.

\subsection{Hubbard and Extended Hubbard Models}

Among the interesting features of the Hubbard and extended Hubbard
models are the variety of phase transitions which can occur as the
density of particles in the ground state changes.  While tuning the
density of particles in the ground state is easily accomplished by
adding a chemical potential to the Hamiltonian, early attempts to
analyze these theories using naive real-space renormalization group
methods ran into problems: projecting onto a small number of states per
block so that the occupation number of each state is a finite integer,
and therefore the density a rational fraction, made it difficult to
achieve a smooth dependence of the density on the chemical potential.
CORE mitigates this problem without having to keep a large number of
states per block:  first, the connected range-$r$ terms are computed by
diagonalizing the full $r$-site Hamiltonian, including the chemical
potential, and so these terms can encode more complicated behavior of
the chemical potential coefficient $\mu$; second, the operator which
measures the density of particles in the ground state as a function of
$\mu$ undergoes a much more complicated evolution than it does in a
naive truncation procedure, evolving connected range-$r$ terms of its
own.  Preliminary computations support this picture but more extensive
computations are needed to fully explore the potential of CORE methods
for this class of problems.

\subsection{Connection to Perturbation Theory}
\label{sec:connect}

In this section, we discuss the way in which one could establish
the relationship between the CORE approach and the familiar
perturbative renormalization group in the weak-coupling limit.  To
illustrate this connection, consider adding a $\lambda\phi^4$
interaction to the scalar field theory Hamiltonian given in
Eq.~\ref{bosefreefield}, where the $\lambda$ coupling is small, and
again apply the CORE procedure outlined in Sec.~\ref{freebose}.  It is
a straightforward exercise to include the $\phi^4$ term and
perturbatively compute the CORE transformation associated with the
two-site, infinite-state truncation procedure.

We begin with the same truncation procedure defined for the free-field
case and keep the same tower of oscillator states for each two-site
block.  A new feature is that we must now compute $S$ and $H_{\cal T}$
even for the range-1 terms because the retained states contract onto
states which are different from the two-site, free-field
eigenstates and the eigenenergies corresponding to these states are
also changed from their free-field values.   A direct consequence of
this is that the new range-1 connected part of the renormalized
Hamiltonian contains higher-order polynomials in the fields.
Given a perturbative expression for the connected range-1 terms, we
have to perturbatively solve the four-site problem to compute
$S$ and $H_{\cal T}$ in order to obtain the connected range-2 terms in
the renormalized Hamiltonian.  Once again, we get a set of terms of the
form $\phi^{m}(p)\phi^{n}(p+1)$ which do not correspond to terms in
the original Hamiltonian.  Longer-range connected terms are computed in
the same manner.  Since the zero-coupling limit of this
procedure builds up a finite-range expansion of the free-field theory,
one should be able to make this perturbative expansion match up with
more familiar renormalization group computations.

There is a simple way to modify the procedure just outlined so as to
automatically resum the perturbative expansion of the renormalized
Hamiltonian to very high order in the coupling.  One virtue of this
modified approach is that it guarantees that the ground-state energy
density will behave as $\lambda^{1/3}$ for large couplings.  The basic
idea is to change the definitions of $\gamma_{\pm}$ in
Eq.~\ref{annihcreatphifour} in order to treat them as variational
parameters which depend upon $\mu^2$ and $\lambda$.  To determine their
values, we minimize the expectation value of the two-site Hamiltonian
in the state $\vert\gamma_{+},\gamma_{-}\rangle$ with respect to
$\gamma_{+}$ and $\gamma_{-}$.  Fixing $\gamma_{+}$ and $\gamma_{-}$
in this way, we then rewrite the two-site Hamiltonian in terms of
annihilation and creation operators, normal order the resulting
expression, and do perturbation theory in the non-quadratic terms.
Note that this minimization process guarantees that the state
$\vert\gamma_{+},\gamma_{-}\rangle$ is the lowest-lying eigenstate
of the ``free Hamiltonian'' obtained by keeping the quadratic terms,
including those which come from normal ordering the quartic
self-interaction.  Since $\gamma_{+}$ and $\gamma_{-}$ are
nontrivial functions of $\mu^2$ and $\lambda$, the perturbation
theory just described amounts to an infinite resummation of the
usual expansion.

To compute the range-2 terms in the renormalized Hamiltonian,
solve the four-site free problem but treat the $\gamma$
frequencies as variational parameters determined by minimizing
the expectation value of the Hamiltonian in the ground state of
the oscillators.  This leads to four coupled equations
which can be solved numerically for any value of $\lambda$.  Once
again, normal order the Hamiltonian and treat all terms which are
not quadratic in the ladder operators as perturbations.  The states
obtained by working to finite order in these perturbations are used to
construct $S$.  The computation of higher-range connected terms
proceeds in a similar manner.

\section{Conclusion}
\label{sec:conclude}
The contractor renormalization group, a general method for
solving any Hamiltonian lattice system, was presented.  The CORE
approach is a systematic and nonperturbative procedure for carrying
out real-space renormalization group transformations which
relies on contraction and cluster techniques.  The method
was illustrated using four examples: free scalar field theory
with single-state truncation, the Heisenberg antiferromagnetic spin
chain with two-state truncation, the anisotropic Ising model with
two-state truncation, and free scalar field theory with an
infinite-state truncation scheme.  The use of approximate contractors,
the convergence of the cluster expansion in determining the
renormalized Hamiltonian and the need for summation via the
renormalization group were also discussed.

A particularly exciting feature of the CORE technology is its
ability to treat systems with dynamical fermions, systems which
are difficult to study using stochastic methods.  CORE also makes
possible gauge-invariant renormalization group transformations in
Hamiltonian lattice gauge theory and easily incorporates a
chemical potential.  These features suggest that the CORE
approximation will prove to be a powerful tool in future
applications to the Hubbard and t-J models and lattice
gauge theory with and without fermions.

This work was supported by the U.~S.~DOE, Contract
No.~DE-AC03-76SF00515 and Grant DE-FG03-90ER40546,
and the UK PPARC through grant GR/J 21347.

\appendix
\section{Harmonic Oscillator Identities}
\label{app:sho}

Consider the canonically conjugate operators $x$ and $p$ which
satisfy $[ x, p ] = i$, and introduce the one-parameter family of
annihilation and creation operators
$A_\omega = x\sqrt{\omega/2}+ip/\sqrt{2\omega}$ and
$A^\dagger_\omega = x\sqrt{\omega/2}-ip/\sqrt{2\omega}$
which satisfy $[ A_\omega, A_\omega^\dagger] = 1$.  Note that
$x = ( A^\dagger_\omega + A_\omega )/ \sqrt{2\omega}$ and
$p = i\,(A^\dagger_\omega - A_\omega )\sqrt{\omega/2}$.
Furthermore, define the Hamiltonian
$H_{\omega} = \frac{1}{2} ( p^2 + \omega^2\,x^2 )$.

\begin{quote}
{\bf Theorem A1:} The state $\vert\omega_1\rangle$ defined by
$A_{\omega_1}\vert\omega_1\rangle=0$ and the state $\vert\omega_0\rangle$
defined by $A_{\omega_0}\vert\omega_0\rangle=0$ are related by
\begin{equation}
   \vert\omega_1\rangle =  \left( \frac{4\,\omega_0\,\omega_1}
    {(\omega_0 + \omega_1)^2}\right)^{1/4}
     \exp\left\{\frac{(\omega_0 - \omega_1)}{ 2 (\omega_0 + \omega_1)}
     A_{\omega_0}^{\dagger 2}\right\}
      \,\vert\omega_0\rangle.
   \label{avii}
\end{equation}

{\sl Proof:}
Write $A_{\omega_1}$ in terms of $x$ and $p$, then express $x$
and $p$ in terms of $A_{\omega_0}$ and $A^\dagger_{\omega_0}$ to
show that
 \begin{equation}
      A_{\omega_1}= \gamma_M \, A^\dagger_{\omega_0} +
         \gamma_P \, A_{\omega_0},
   \label{aix}
\end{equation}
where
\begin{eqnarray}
   \gamma_M &=& {\textstyle\frac{1}{2}}
     \,\left(\,\sqrt{\omega_1/\omega_0}
     - \sqrt{\omega_0/\omega_1} \right), \\
   \gamma_P &=& {\textstyle \frac{1}{2}}
   \,\left(\, \sqrt{\omega_1/ \omega_0}
     + \sqrt{\omega_0/ \omega_1} \right).
   \label{ax}
\end{eqnarray}
Now use $[A_{\omega_0}, A_{\omega_0}^{\dagger n}]
 =nA_{\omega_0}^{\dagger n-1}$ to show that
 \begin{equation}
   \left[A_{\omega_0}, \exp\left(\xi A_{\omega_0}^{\dagger 2}\right)
   \right]= 2\xi\ \exp\left(\xi A_{\omega_0}^{\dagger 2}\right)
   A_{\omega_0}^\dagger.
 \end{equation}
Hence,
\begin{equation}
 A_{\omega_1} \exp\left(\xi A_{\omega_0}^{\dagger 2}\right)
 \vert\omega_0\rangle = \left(\gamma_M+2\xi\gamma_P\right)
 \exp\left(\xi A_{\omega_0}^{\dagger 2}\right)
 A_{\omega_0}^\dagger \vert\omega_0\rangle,
\end{equation}
which vanishes if we set $\xi=-\gamma_M/(2\gamma_P)
=(\omega_0-\omega_1)/[2(\omega_0+\omega_1)]$. Since
$A_{\omega_1}\vert\omega_1\rangle=0$ defines $\vert\omega_1\rangle$,
then clearly
\begin{equation}
N_{\omega_1}
 \exp\left\{-\gamma_M A_{\omega_0}^{\dagger 2}/(2\gamma_P)\right\}
  \vert\omega_0\rangle = \vert\omega_1\rangle.
\end{equation}
Requiring $\langle\omega_1\vert\omega_1\rangle=1$
and using $\langle\omega_0 \vert A_{\omega_0}^n A_{\omega_0}^{\dagger
 m}\vert\omega_0\rangle=\delta^{mn}n!$, we have
 \begin{equation}
    N_{\omega_1}^{-2} = \sum_{n=0}^\infty
      \frac{(2n)!}{(n!)^2}\xi^{2n}
      = (1 - 4\xi^2)^{-1/2},
   \label{axix}
 \end{equation}
so $ N_{\omega_1} = [ 4\,\omega_1\,\omega_0/
            (\omega_1 + \omega_0)^2 ]^{1/4}$.
\end{quote}

Given this result, we can now easily show that applying
$e^{-t H_{\omega_0}}$ to an arbitrary Gaussian wavefunction
produces a new Gaussian wavefunction of a different frequency;
as $t\rightarrow\infty$, the new frequency tends to $\omega_0$.

\begin{quote}
{\bf Theorem A2:}
Let $\vert\omega_1\rangle$ be the simple harmonic oscillator
ground state defined by $A_{\omega_1}\vert\omega_1\rangle=0$ and
$H_{\omega_0}$ be the Hamiltonian for a simple harmonic oscillator
of frequency $\omega_0$.  Then
\begin{equation}
e^{-tH_{\omega_0}}\vert\omega_1\rangle=A(t)\vert\omega(t)\rangle,
\label{shoevolve}
\end{equation}
where $\vert\omega(t)\rangle$ is the ground state of a simple harmonic
oscillator of frequency $\omega(t)$, and
\begin{eqnarray}
   A(t) &=& \left(\frac{\omega_1}{\omega(t)}\right)^{\frac{1}{4}}
     \left(\frac{\omega(t) + \omega_0}{\omega_1
     + \omega_0 } \right)^{\frac{1}{2}}  \,e^{-t\,\omega_0/2},
     \label{axxix}\\
   \omega(t) &=& \omega_0\, \left(\frac{  \omega_0 + \omega_1
   - e^{-2\,t\,\omega_0}( \omega_0 - \omega_1)}{
        \omega_0+\omega_1 + e^{-2\,t\,\omega_0}(\omega_0 -
         \omega_1 )}  \right).
   \label{axxv}
 \end{eqnarray}

{\sl Proof:}
Using $e^{-t\,H_{\omega_0}}\,A_{\omega_0}^{\dagger m}
\,e^{t\,H_{\omega_0}} =
  e^{-m\,t\,\omega_0} \,A_{\omega_0}^{\dagger m}$, one sees that
\begin{equation}
e^{-t\,H_{\omega_0}}\,e^{\xi A^{\dagger 2}_{\omega_0}}
\,e^{t\,H_{\omega_0}} =  e^{\xi^\prime A^{\dagger 2}_{\omega_0}},
\end{equation}
where $\xi^\prime=e^{-2t\omega_0}\xi$.   Using
$e^{-t\,H_{\omega_0}}\,\vert\omega_0\rangle = e^{-t \omega_0 /2}
 \vert\omega_0\rangle$ and Eq.~\ref{avii}, one finds that
 \begin{equation}
   e^{-t\,H_{\omega_0}}\,\vert\omega_1\rangle = N_{\omega_1}
   e^{-t\omega_0/2}
   \exp\left\{\frac{(\omega_0 - \omega_1)}{ 2 (\omega_0 + \omega_1)}
   e^{-2t\omega_0} A_{\omega_0}^{\dagger 2}\right\}
   \,\vert\omega_0\rangle.
   \label{axxo}
\end{equation}
If we set
 \begin{equation}
   {(\omega_0 - \omega(t) ) \over 2\,(\omega(t) + \omega_0) } =
      e^{-2\,t\,\omega_0}\,{(\omega_0 - \omega_1) \over 2\,(\omega_0
       + \omega_1) },
   \label{axxiv}
 \end{equation}
then we can identify the state on the right-hand side of
Eq.~\ref{axxo} with the Gaussian wavefunction $\vert\omega(t)\rangle$.
Solving Eq.~\ref{axxiv} for $\omega(t)$ yields the result given
in Eq.~\ref{axxv}.  Note that $\omega(0)=\omega_1$ and
$\omega(t\rightarrow\infty)=\omega_0$.
The multiplicative factor $A(t)$ is then given by
$A(t)=(N_{\omega_1}/N_{\omega(t)})e^{-t\omega_0/2}$ which simplifies
to the result shown in Eq.~\ref{axxix}.
\end{quote}

\newpage
\epsfverbosetrue

\epsfverbosetrue

\epsfverbosetrue

\epsfverbosetrue

\epsfverbosetrue

\epsfverbosetrue

\epsfverbosetrue

\epsfverbosetrue

\epsfverbosetrue

\begin{table}
\caption[bose]{Convergence of the partial sums $\epsilon_n$ in the
cluster expansion of the free massless scalar-field vacuum energy
density as a function of the range $n$.  The energies are
CORE estimates from a single-state per site truncation
algorithm, and the errors are the differences of these
estimates from the exact energy density 0.636619772.}
\label{fftable}
\begin{center}
\begin{tabular}{rlc}
 $n$ &  \multicolumn{1}{c}{$\epsilon_n$} &   Error \\
   1 &   0.707107  &   0.070487  \\
   2 &   0.658919  &   0.022299  \\
   3 &   0.647644  &   0.011025  \\
   4 &   0.643206  &   0.006586  \\
   5 &   0.641001  &   0.004382  \\
   6 &   0.639746  &   0.003126  \\
   7 &   0.638962  &   0.002343  \\
   8 &   0.638441  &   0.001821  \\
   9 &   0.638076  &   0.001456  \\
  10 &   0.637811  &   0.001191  \\
  20 &   0.636932  &   0.000312  \\
  30 &   0.636761  &   0.000141  \\
  40 &   0.636700  &   0.000080
\end{tabular}
\end{center}
\end{table}
\newpage

\begin{table}
\caption[haf]{Comparison of some of the couplings $\alpha_s^{(r)}$
in the renormalized Hamiltonian of a free scalar field theory for
$r=2,3,4,\infty$, where $r$ is the cluster-expansion truncation
order.  Note that only $\alpha_0$ and $\alpha_1$ are
defined for the case $r=2$.
}
\label{alphatable}
\begin{center}
\begin{tabular}{crrrr}
$\alpha_s^{(r)}$ & \multicolumn{1}{c}{$r=2$} &
 \multicolumn{1}{c}{$r=3$} & \multicolumn{1}{c}{$r=4$} &
 \multicolumn{1}{c}{$r=\infty$} \\
$\alpha_0$ & $ 0.381966$ & $ 0.371054$ & $ 0.367594$ & $ 0.363380$ \\
$\alpha_1$ & $-0.500000$ & $-0.451922$ & $-0.438360$ & $-0.424413$ \\
$\alpha_2$ &             & $ 0.104212$ & $ 0.098137$ & $ 0.084883$ \\
$\alpha_3$ &             &             & $-0.041400$ & $-0.036378$
\end{tabular}
\end{center}
\end{table}
\begin{table}
\caption[haf]{Comparison of truncated-cluster ${\cal E}_r$ and
finite-volume $E_r/r$ Heisenberg antiferromagnet
ground-state energy densities for range $r$.}
\label{haftable}
\begin{center}
\begin{tabular}{ccc}
$r$ & ${\cal E}_r$ & $E_r/r$  \\
  2 & $-0.3750000$ & $-0.3750000$ \\
  4 & $-0.4330125$ & $-0.4040063$ \\
  6 & $-0.4387760$ & $-0.4155962$ \\
  8 & $-0.4406775$ & $-0.4218665$
\end{tabular}
\end{center}
\end{table}

\end{document}